%% file: fabula_arxiv.tex
\documentclass[11pt, a4paper, logo, copyright, numbering]{googledeepmind}

\usepackage[authoryear, sort&compress, round]{natbib}

\usepackage{makecell}
\usepackage{xcolor}

\begin{document}

\title{Fabula: Building a Narrative Storytelling Sidekick with the Writers' Community}

\correspondingauthor{Piotr Mirowski piotrmirowski@google.com}

\reportnumber{} 

\renewcommand{\today}{}

\author[1,*]{Piotr Mirowski}
\author[1,*]{Ben Wedin}
\author[1, *]{Reinald Kim Amplayo}
\author[1]{Rich Galt}
\author[1]{Duncan Williams}
\author[1]{Rida Qadri}
\author[1]{Jaume Sanchez-Elias}
\author[1]{Erin Drake-Kajioka}
\author[1]{Sian Gooding}
\author[1]{Lucia Lopez-Rivilla}
\author[1]{Joao G. M. Araujo}
\author[2]{Lion Schulz}
\author[1]{Satinder Baveja}
\author[1]{Shakir Mohamed}
\author[1]{Edward Grefenstette}
\author[1]{Laura Rimell}
\author[1,*]{Richard Evans}

\affil[*]{Equal contributions}
\affil[1]{Google DeepMind}
\affil[2]{Bertelsmann}

\newcommand{\dramabox}[0]{Fabula}
\newcommand{\dramaboxlink}[0]{\url{https://deepmind.google.com/frontiers/fabula}}

\newcommand{\pquote}[1]{{``\textit{#1}''}}
\newcommand{\nquote}[2]{{``\textit{#1}'' (#2)}}

\definecolor{todo}{rgb}{1,0,0}
\newcommand{\todo}[1]{{\color{todo} #1}}

\newcommand{\Description}[1]{}

\begin{abstract}
\input{paper_abstract}
\end{abstract}

\keywords{storytelling, screenwriting, playwriting, narratology, creativity, participatory design, large language models, user interface, cultural AI evaluations, human-computer interaction}

\maketitle

\begin{figure}
  \includegraphics[width=0.49\textwidth]{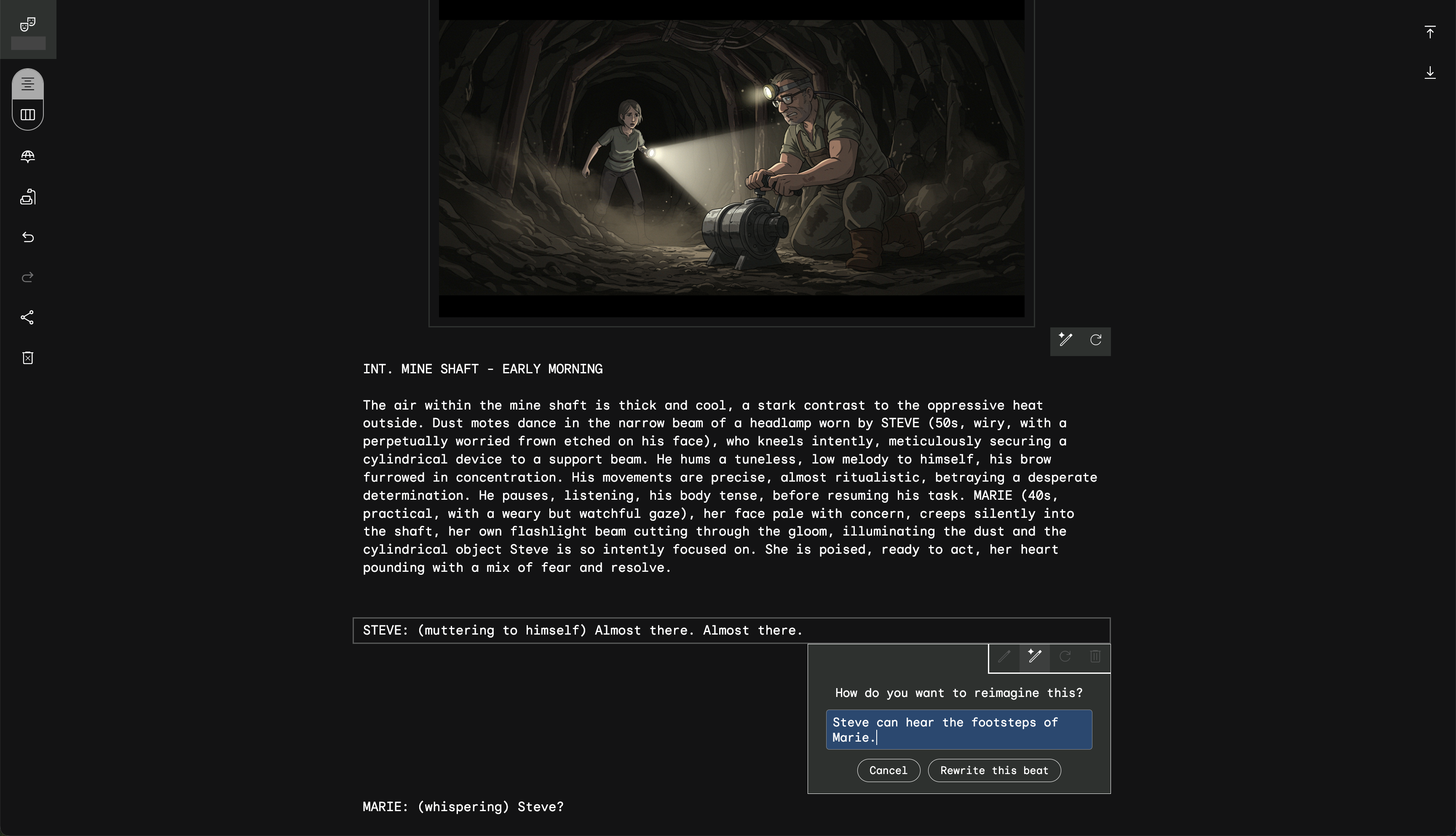}
  \includegraphics[width=0.49\textwidth]{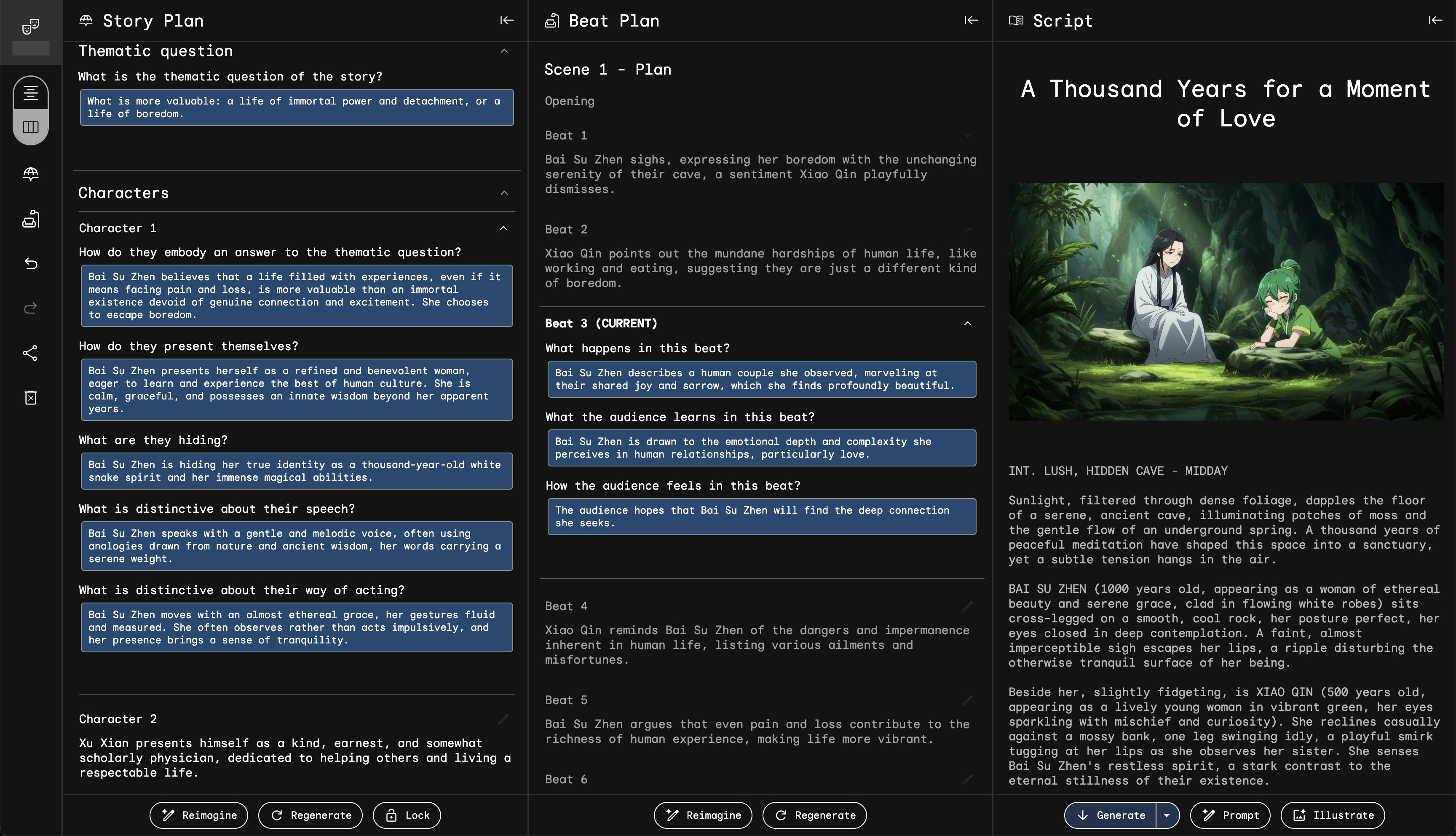}
  \caption{\dramabox~user interface Left: Gardener view. Right: Architect views. Boxes in blue are editable.}
  \Description{Two screen captures showing the user interface for \dramabox. The left one shows the Gardener mode (single column with script), the right one shows the Architect mode (three columns, with story plan, scene plan and script). The images show that the script consist of dialogue, descriptions and image illustrations. The images show blue boxes near certain elements of the interface, corresponding to current story edits.}
  \label{fig:teaser}
\end{figure}

\section{Introduction}
\input{paper_introduction}

\section{Related Work}
\input{paper_related_work}

\section{Methods}
\input{paper_methods}

\section{Design Choice 1: Self-Improving Story Generators}
\input{paper_design_quality}

\section{Design Choice 2: Gardener or Architect UI Layout Instead of Chatbot}
\input{paper_design_architect}

\section{Design Choice 3: Hierarchical Scene-Beat Story Plans that Accommodate Culturally-Situated Narratives}
\input{paper_design_hierarchical}

\section{Design Choice 4: Convergent Iteration for Creative Writing}
\input{paper_design_creativity}

\section{Discussion}
\input{paper_discussion}

\section*{Conclusion}
\input{paper_conclusion}

\input{acknowledgements}

\bibliographystyle{ACM-Reference-Format}
\bibliography{fabula_arxiv}

\appendix

\section{Ethics Considerations}
\input{appendix_ethics}


\section{Experimental Protocols}
\input{appendix_user_survey}
\input{appendix_design_sessions}
\input{appendix_creativity_survey}
\input{appendix_writer_sessions}

\end{document}

%% file: paper_abstract.tex
We design and evaluate \dramabox, an interactive app for fiction writers. \dramabox~uses detailed narrative plans informed by narratological theory. Stories are structured hierarchically into scenes and beats that can be (re)generated and revised at script and story plan level.
Using participatory AI, we critically evaluate and improve \dramabox~with casual and published writers, via design interviews and writing sessions with 42 experts, and large-scale testing.
We interrogate our design choices: (1) whether a language model-based auto-evaluator, optimized on human experts’ preferences, can improve story quality, (2) whether users want UI that exposes the detailed narrative plan alongside the story script, (3) to what extent our narratology assumptions fit localised storytelling traditions and serve screenwriters or playwrights, and (4) whether convergent iteration over the story plan supports writers' creativity.
Building on critical feedback and concerns, we use \dramabox~as a cultural probe, and identify potentials for education and interactive storytelling.

%% file: paper_introduction.tex
AI-generated media has the potential to transform culture and entertainment, as artists are trying AI tools to explore new interactive experiences or forms of storytelling. AI has also become a hotspot of tensions and questions around creativity, agency and livelihoods \citep{gero2025creative,li2024value,tang2025understanding}. The writers' community itself is seeing a fissure with some award-winning writers using LLM based co-writing tools \citep{goldberg2024playwright,ai2021tang}, and others eschewing them completely for ethical reasons \citep{chambers2025ai}.

\subsection{Participatory development of an AI tool for creativity support and co-writing stories}

There is an increasing desire to bring in writers' and other creatives' voices into the design of AI that aligns with creative values and desires of artists \cite{kim2024authors,lee2024design,biermann2022tool}, but most participatory frameworks \citep{birhane2022power} treat user feedback merely to improve tools---what \citet{sloane2022participation} has called \pquote{participation washing}---rather than as a way to negotiate fundamental, potentially unresolvable, values \cite{delgado2023participatory}. In this paper, instead of using AI software as a final product for feedback, we use prototyping as inquiry \citep{schon1992designing,kannabiran2020prototypes} to understand the clash between our design choices and the needs of the creative writing community, which we try to reconcile within the constraints of working in a research and development lab. 

Specifically, we introduce \emph{\bf \dramabox}, an interactive app for supporting fiction writers\footnote{\dramabox~is available to writers who make an access request at \dramaboxlink. \dramabox~data governance is summarised in Appendix \ref{subsec:appendix-data}}. \dramabox~presents a story alongside its detailed narrative plans, structured hierarchically into scenes and beats. The story can be written, generated, regenerated, and revised at script-, scene- and and story plan-level. \dramabox~uses detailed story plans (so-called \emph{Drama Managers}) that mirror classical narratology models used in screenwriting and playwriting.

We use \dramabox~as a ``lightning rod'' or \emph{cultural probe} \citep{gaver1999design} in adversarial design \citep{disalvo2015adversarial} to surface and anchor the abstract, often invisible ideological conflicts between tool developers and artistic practitioners. By iteratively ``moving'' the object, altering its affordances, constraints, and user interface (UI), we transform the design process and development cycle into a diagnostic method. We observe how specific features become sites of contestation and negotiation of how to best turn \dramabox~into a creativity support tool for established writers or a writing mentor for novice writers. Through this process, \dramabox~becomes a boundary object - a form of critical design \citep{bardzell2012critical} - that allows us to have rooted discussions about the future of creativity with AI. The limitations of \dramabox~and LLMs for writing are just as important as their features because they define the ``boundaries'' of what writers are willing to cede to AI.

The participatory involvement consists of extensive internal testing and evaluation with hundreds of colleagues, as well as collaboration with the writer communities: design interviews with 18 writing experts, writing sessions with 25 writers, and a limited public release to over 250 writers. We followed a process of iterative development where we implemented feedback from one phase in preparation for the next evaluation phase.

\subsection{Paper structure and developer hypotheses and design choices}

Section \ref{sec:literature_review} reviews related work and Section \ref{sec:methods} presents \dramabox, namely the Drama Managers, the app and its user interface, and the various types of computer-human interaction evaluations. We then proceed to evaluate four design choices and developer hypotheses:
\begin{enumerate}
    \item Can self-improving story generators help us build a writing tool that supports writers with high-quality narratives and illustrations (Section \ref{sec:quality})?
    \item Can we depart from chat-based AI writing tools to create custom UIs that support both \emph{Gardeners} and \emph{Architects} writer archetypes (Section \ref{sec:gardeners-architects})?
    \item Can we rely on hierarchical Scene-Beat story plans to structure the story while accommodating culturally-situated narratives (Section \ref{sec:hierarchy})?
    \item Can the \emph{Convergent Iteration} framework support creative writing (Section \ref{sec:creativity})?
\end{enumerate}
The broad scope of evaluation naturally arises from our undertaking of designing a complex and exploratory tool released to thousands of testers with diverse expertises.
Section \ref{sec:discussion} summarises learnings from our research and avenues in which we synthesize our learnings to specialise the writing tool for specific usages.

%% file: paper_related_work.tex
\label{sec:literature_review}

\subsection{Diverse storytelling traditions}
\label{subsec:storytelling}
Narrative theories have been extensively studied in such works as \emph{Story: Substance, Structure, Style and the Principles of Screenwriting} by \citep{mckee1997story} and \emph{Into the Woods: How Stories Work and Why We Tell Them} by \citep{yorke2013into}, which can be seen as manuals of how to write stories, and \emph{The Classical Plot and the Invention of the Western Narrative} by \citep{lowe2000classical}, which analyses the plot from the perspective of cognitive science. These experts often recommend hierarchical plans in which stories are divided into scenes, and scenes are divided into beats. Starting from \citet{aristotle350BCpoetics}'s three-act structure, many dramatic structures have been devised~\citep{polti1917thirty}, including popular variations on the \emph{Freytag's Pyramid}~\citep{freytag1894technik} or the Hero's Journey and \emph{Monomyth} \citep{campbell2008hero,vogler2007writer}. Many such narrative structure are however characteristically Western~\citep{johnstone2005discourse, de2015discourse} and alternative story shapes, or ``story grammars''~\citep{rumelhart1980evaluating} are possible~\citep{becker1979text, chafe1980pear}, including the four-part structure of \emph{Kishōtenketsu} that is prevalent in Chinese, Japanese or Korean storytelling.

\subsection{State-of-the-art in story generation}
While current LLMs are uncannily capable at generating long-form text, zero-shot LLM outputs have been perceived as inferior to human-generated writing. Detailed analysis of LLMs for co-writing identified repetitiveness, reliance on clichés and tropes, the lack of nuance, subtext and symbolism, and overly moralistic and predictable endings as some of the categories of concerns from human writers \citep{chakrabarty2024creativity}. Comparisons of LLM-generated passages and human passages demonstrated that as of late 2023, LLMs were not capable of evaluating writing according to tailored tests of creative writing \citep{chakrabarty2024art}.

As new LLM models are released, their apparent creative writing skills are improving. In one recent, thought-provoking experiment by \citet{chakrabarty2025readers}, state-of-the-art LLMs fine-tuned on works of specific writers succeeded in generating short prose in the style of those artists according to MFA-level judges in blind experiments\footnote{Arguably, style imitation is perceived as a contrived task by writers \citep{porquet2025copying} who primarily aim at finding their own voice and expressing their own stories.}.

\subsection{AI writing tools for storytellers}

A few co-creation apps provide writers with highly immersive and interactive interfaces that lets them generate the world of their story as they write, notably \emph{AI Dungeon} \citep{aidungeon2019} and \emph{Lore Machine} \citep{loremachine2024}. They are however geared towards game-like experiences in a limited set of predefined story worlds, and do not allow to view the proposed story's arc or to edit previous parts of the story other than through reconstructing the story piece by piece.

Other AI writing support apps enable the writer to plan their story ahead, by creating structured and detailed blueprints for their story, for example \emph{Sudowrite Muse} \citep{sudowrite2021} and \emph{SAGA} \citep{saga2024}).

\subsection{Creativity Support Systems for storytelling}

AI-powered Creativity Support Systems \citep{massetti1996empirical,abraham1994computer} increasingly focus on high-level co-creation, and researchers try to document and quantify the creative activity traces that results from human-computer interaction in a creative context \citep{kreminski2026herding,hammad2026tracing}.

Early research established the potential for LLMs to aid in story world-building and structure: \citet{ippolito2022creative} explored the human-AI collaborative editor \emph{Wordcraft} as a tool for prose, while \emph{Dramatron} \citep{mirowski2023cowriting} pioneered and investigated the use of hierarchical prompt-chaining --- starting from a logline to generate characters, plot, and dialogue --- specifically for screenwriting and playwriting. \emph{Sparks} \citep{gero2022sparks} further illustrated how LLMs can provide targeted inspiration during the science writing process by generating diverse conceptual starting points. Further studies have since been conducted with professional writers using AI writing tools like \emph{SceneCraft} \citep{kumaran2023scenecraft}, \emph{CharacterMeet} \citep{qin2024charactermeet}, \emph{Drama Llama} \citep{sun2025drama} or \emph{Dramamancer} \citep{wang2025dramamancer}, to name a few, and investigated node-based story structures \citep{wen2026garden,niu2026storycomposerai}, narrative blocks \citep{sillano2026text}, and story remixing \citep{zhang2026narrix,ma2026narrativeloom}.

In this paper, we engage in co-design of the generative AI tool, similarly to work done with filmmakers working with developers of generative video tools \citep{mathewson2025behind} or visual artists exploring ``hacks'' with image generators \citep{qadri2025ai}.

%% file: paper_methods.tex
\label{sec:methods}

This section describes the LLM-based agent (so-called \emph{Drama Manager}, Section \ref{subsec:drama-managers}), the \dramabox~app (Section \ref{subsec:dramabox-app}), as well as its development cycle (Section \ref{subsec:cycle}) that tightly incorporates human-computer interaction methods (Section \ref{subsec:hci-methods}).

\subsection{Drama Managers}
\label{subsec:drama-managers}

\subsubsection{Hierarchical Scene-Beat modeling of stories with a story plan and scene plans.}
Robert McKee's \emph{Story} \citep{mckee1997story} defines a scene as action \pquote{that turns the value-charged condition of a character's life on at least one value with a degree of perceptible significance.}
He describes the beat as: \pquote{the smallest element of structure [...] an exchange of behavior in action/reaction.}

In \dramabox, we adopt this hierarchical view. A story is represented as a sequence of scenes, and each scene is a sequence of beats. A \dramabox~\emph{beat} is a section of the script of a few lines in which something dramatically significant happens. This maps to Konstantin Stanislavsky's \emph{bit} or \emph{unit}: a section of the script where character objectives remain constant \citep{benedetti2005stanislavski}.

Following narrative craft guidelines, we define beat changes as being triggered by: changes in the physical world (e.g., entrances or exits), location shifts, topic changes in dialogue, shifts in emotional states or power dynamics, or changes in character objectives.
Each beat is designed to be short (typically up to 10 lines) but can contain multiple dialogues or actions.


\subsubsection{Drama Managers.}
To organize story structures and generate stories, \dramabox~delegates planning and text generation to a \emph{drama manager}. The base drama manager is an abstract Python interface defining the contract for story generation. It encapsulates methods for:
\begin{itemize}
    \item \textbf{Plan Creation}: Generating the initial hierarchical plan.
    \item \textbf{Content Generation}: Producing the actual script lines for beats (`next\_lines`).
    \item \textbf{Plan Evolution}: Updating plans dynamically based on user feedback (e.g., retrying beats, extending scenes, or modifying specific plan elements).
\end{itemize}
All of these functions work by invoking LLMs (Gemini 2.5 Flash \citep{comanici2025gemini} at the time of the study\footnote{As of February 2026, \dramabox~relies on Gemini 3 Flash and Gemini 3.1 Pro.}) with structured outputs to ensure consistency, and multiple times with a large range of different prompts. All drama managers operate on three levels of abstraction: the story plan, the beat plan, and the script.

Each drama manager allows modifying the story at each layer of abstraction. One can modify the story plan at high-level (e.g., ``X vanishes in scene 4”) or by fine-grained edits to particular aspects of the story plan (e.g., a character's backstory).
Similarly, one can modify the beat plan at high-level 
or by fine-grained edits (e.g., editing what happens in beat 6, scene 4). One can also modify the story text itself by high-level suggestions, or by fine-grained textual edits.

We implemented 26 different drama managers to explore the tradeoff between story quality and system responsiveness (due to LLM latency). Some models focus on particular narrative theories \citep{mckee1997story,lowe2000classical,yorke2013into}, and some are amalgamations of different theories. As discussed in Section \ref{subsec:self-play}, this wide variety allows us to test empirically which aspects of narrative planning are most effective for improving story quality.

\subsubsection{Drama Manager design}
Our approach to drama manager design is iterative: each new capability is developed in response to specific shortcomings surfaced through feedback from writers, or by our automated evaluation system (see Section \ref{subsec:self-play}). When a problem is identified (such as \emph{cliché} plot or characters with unbelievable motivations), we consult established theories in cognitive science and narrative craft and add a modular component to the drama manager.

Certain writers (see Section \ref{sec:hierarchy}) critiqued the existence of universal story structures that can be deployed across diverse writing cultures. We acknowledge the limitations of our approach (iterating on the design of a single drama manager) and propose, for further work, to use a selected drama manager as a point of departure for further refinements and adaptation.

\subsubsection{Story- and scene-level plans.}
The foundational component to ensure story coherence and structure, represents the story as a two-level hierarchical plan with the story plan and the beat plan. The higher-level story plan is composed of individual scenes, where each scene is defined by: \emph{what the audience learns}, \emph{how the audience feels}, \emph{what happens}, \emph{the social situation in which the scene is embedded}, and \emph{the location and time}. The lower-level beat plan breaks down each scene into a sequence of beats. Each beat contains \emph{what the audience learns}, \emph{how the audience feels}, and \emph{what happens}.

This method of story planning is inspired by Nick Lowe's cognitive model of the reader who continuously updates their mental model and affective state as they process the story \citep{lowe2000classical} - a design choice that works for popular storytelling but may be at odds with contemporary theatre (see Section \ref{sec:hierarchy}).

\subsubsection{Elaborate narrative elements.}
This component enriches the story plan with fleshed-out characters, complex relationships, and thematic resonance, inspired by ideas in Yorke's \emph{Into the Woods} \citep{yorke2013into}. To steer clear of clichés and plagiarism, the module first prompts the LLM to generate a list of \emph{well-known stories to avoid}.

Central to this component is the \emph{existential question} (e.g., ``Is success more important than authenticity?''), which serves to unify the narrative. Each character is then developed relative to this question:
\begin{itemize}
    \item \textbf{Embodiment}: How the character's actions implicitly answer the existential question.
    \item \textbf{Presentation vs. Reality}: Contrasting the public face the character presents (\emph{how they present themselves}) with their inner secrets (\emph{what they are hiding}). This design focuses on contrasts between outward presentation and unarticulated forces beneath the surface \citep{goffman1949presentation}.
    \item \textbf{Distinctive Traits}: Specific details about their unique way of speaking and acting to ensure memorable characterization.
\end{itemize}

Similarly, relationships are not merely labeled (e.g., ``friends'') but detailed with \emph{what the relationship is really like} (e.g., power imbalances), adding depth beyond structural roles.

\subsubsection{Audience and narrator goals.}
This component introduces a more deliberate approach to story construction by aligning the narrative with reader expectations and authorial intent. We prompt the LLM to generate:
\begin{itemize}
    \item \textbf{Audience Goals}: What the audience wants to see happen, find out, or feel (e.g., ``The audience wants to understand the source of the conflict'').
    \item \textbf{Narrator Goals}: Explicit extra-diegetic instructions for the narrator to satisfy those audience goals (e.g., ``Gradually reveal the protagonist's flaw'').
    \item \textbf{World-building Questions and Answers}: Crucial uncertainties that must be resolved (e.g., ``What is the secret weapon?'') and their concrete answers before detailed scene planning begins.
\end{itemize}

\subsubsection{Narrative desiderata.}
To actively capture and hold the reader's attention, this module forces the LLM to explicitly plan for a set of narrative desiderata before outlining scenes. Inspired by \citep{mckee1997story,lowe2000classical,yorke2013into}, these include:
\begin{itemize}
    \item \textbf{Thematic Unity}: Ensuring characters' actions implicitly explore the existential question.
    \item \textbf{Suspense}: Highlighting areas of uncertainty in the reader's mental model.
    \item \textbf{Surprise}: Engineering moments that force the reader to revise their understanding, often achieved via internal, interpersonal, or extra-personal conflict.
    \item \textbf{Escalation}: Rising tension and stakes, culminating in a crisis.
    \item \textbf{Closure}: Ensuring endgame conditions are clear early on.
    \item \textbf{Intelligibility}: Verifying that character goals, decisions, and surprises are retrospectively logical.
    \item \textbf{Emotional Range}: Deliberately alternating between positive/negative emotions and tension/release.
\end{itemize}


\subsubsection{Character motivations.}

To ensure believable action and meaningful conflict, this component adds psychological depth borrowing from Stanislavsky's acting theory \citep{benedetti2005stanislavski} and interactive fiction with social simulations \citep{evans2013versu}.
For each significant character in a scene outline, the LLM must explicitly define:
\begin{itemize}
    \item \textbf{Objective}: What the character is actively trying to achieve in this scene.
    \item \textbf{Stakes}: What is at risk if they fail. This must be significant to the character and intelligible to the reader.
    \item \textbf{Obstacle}: What prevents the goal from being easily achieved, ensuring dramatic tension.
\end{itemize}
This forces the generation of scenes driven by character intent rather than arbitrary plot movements.



\subsubsection{Self-critique and Search.}

To elevate the quality beyond a first draft, advanced drama managers use a generate-evaluate-select search loop. This component mimics the process of revision and self-critique (chain-of-thought \citep{wei2022chain}) to refine the story's quality but is computationally expensive. At any planning or generation stage (e.g., story outline, beat sequence, or script dialogue), the module can:
\begin{itemize}
    \item \textbf{Diverse Sampling}: Generate $n$ distinct candidates by iteratively prompting the LLM to avoid previous suggestions.
    \item \textbf{Concurrent Sampling}: Generate $n$ candidates in parallel when diversity is less critical than exploration.
    \item \textbf{Critic Evaluation}: Use a dedicated LLM prompt acting as an expert editor to list pros and cons for each candidate relative to the task prompt.
    \item \textbf{Selection}: Pick the highest-rated candidate based on the critic's chain-of-thought selection argument.
\end{itemize}
While this search process yields significantly higher quality narratives according to our evaluations, it incurs substantially higher latency due to the multiple generation and critique rounds.



\subsection{\dramabox~app}
\label{subsec:dramabox-app}

\begin{figure}
    \centering
    \includegraphics[width=0.8\linewidth]{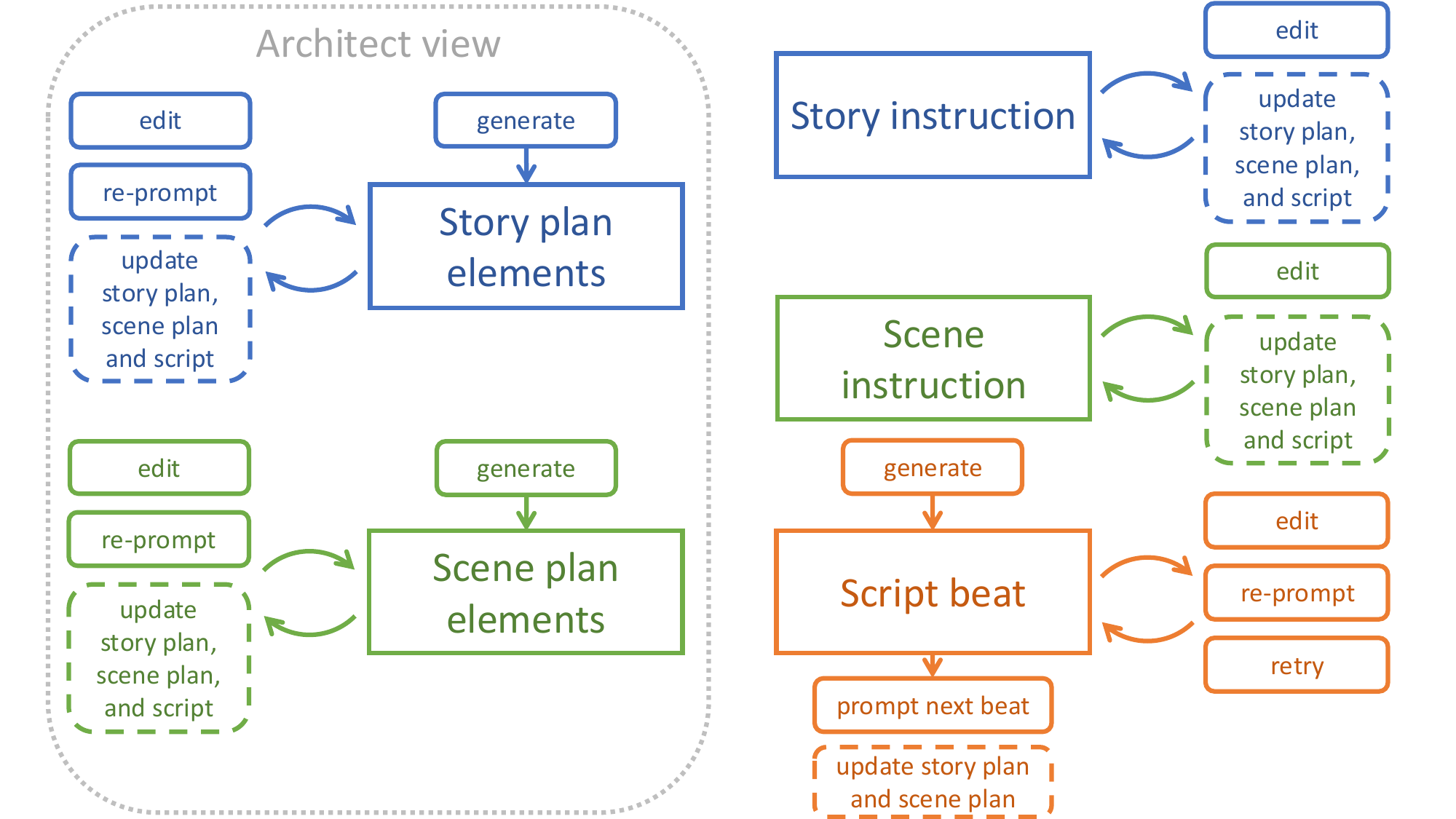}
    \caption{Schematic of interactions within \dramabox.}
    \Description{Schematic showing the types of interactions in the app. On the left, the Architect View box with dashed lines shows two groups of boxes: "story plan elements" and "scene plan elements", each with circular arrows to "edit", "re-prompt" and "update story plan, scene plan and script", and each with incoming arrows from "generate". On the right, there are 3 boxes: "story instruction", "scene instruction" and "script beat". Story instruction and scene instruction have circular arrows to "edit" and "update story plan, scene plan and script". Script beat has an incoming arrow from "generate", circular arrows to "edit", "re-prompt" and "retry" and outgoing arrows to "prompt next beat" and "update story plan, scene plan and script".}
    \label{fig:app-interactions}
\end{figure}

\subsubsection{Main functionalities of the \dramabox~app.}
\dramabox~is a web-based Angular app with a Python backend that queries an LLM (out-of-the-box Gemini 2.5 \citep{comanici2025gemini} or more recent models). The writer interacts with drama managers via an app with a custom user interface (Fig. \ref{fig:teaser}) with two views: the \emph{Gardener} view (showing the script and story instructions only) and the \emph{Architect} view (displaying the story plan alongside the script). The app's internal state is stored in a project snapshot, a JSON object with answers to drama manager questions in previous Section \ref{subsec:drama-managers}, the script so far, and user instructions. For each new user instruction (e.g., ``generate a story plan given a story prompt'', ``revise story plan'', ``manually write next beat'' or ``generate an illustration''), a new story snapshot is created (which enables both implementing an ``undo'' function and retrospectively analysing the story writing process). A schematic view of interactions within \dramabox is shown on Figure \ref{fig:app-interactions}.

In addition to text, \dramabox~enables users to illustrate stories by generating images, and automatically illustrates each new scene. To inspire the writer, \dramabox~generates 3 suggestions\footnote{We ensure diversity of suggestions by vector embedding the corresponding strings, storing past suggestions in a database, and discarding suggestions whose cosine simularity of embeddings are too close to existing suggestions.} of story or scene instructions (synopsis).

\subsubsection{Parsing existing scripts for continuation and alternative generation.}
During internal evaluations and design sessions with published writers (Section \ref{subsubsec:upload}), it emerged that writers wanted \dramabox~to help them complete a partially-written story (e.g., TV screenwriters ideating on new episodes); this could enable generate alternative versions of an existing script, or serve to analyze that existing script using narratological decomposition. During upload, the script is parsed into a sequence of scenes, each of which is a sequence of beats, and \dramabox~constructs the drama manager, story plan and beat plan based on the script. The writer can then evaluate multiple story branches by editing scene plans and generating alternative continuations of their script.


\subsection{Iterative development cycle}
\label{subsec:cycle}

\begin{figure}
    \centering
    \includegraphics[width=0.67\linewidth]{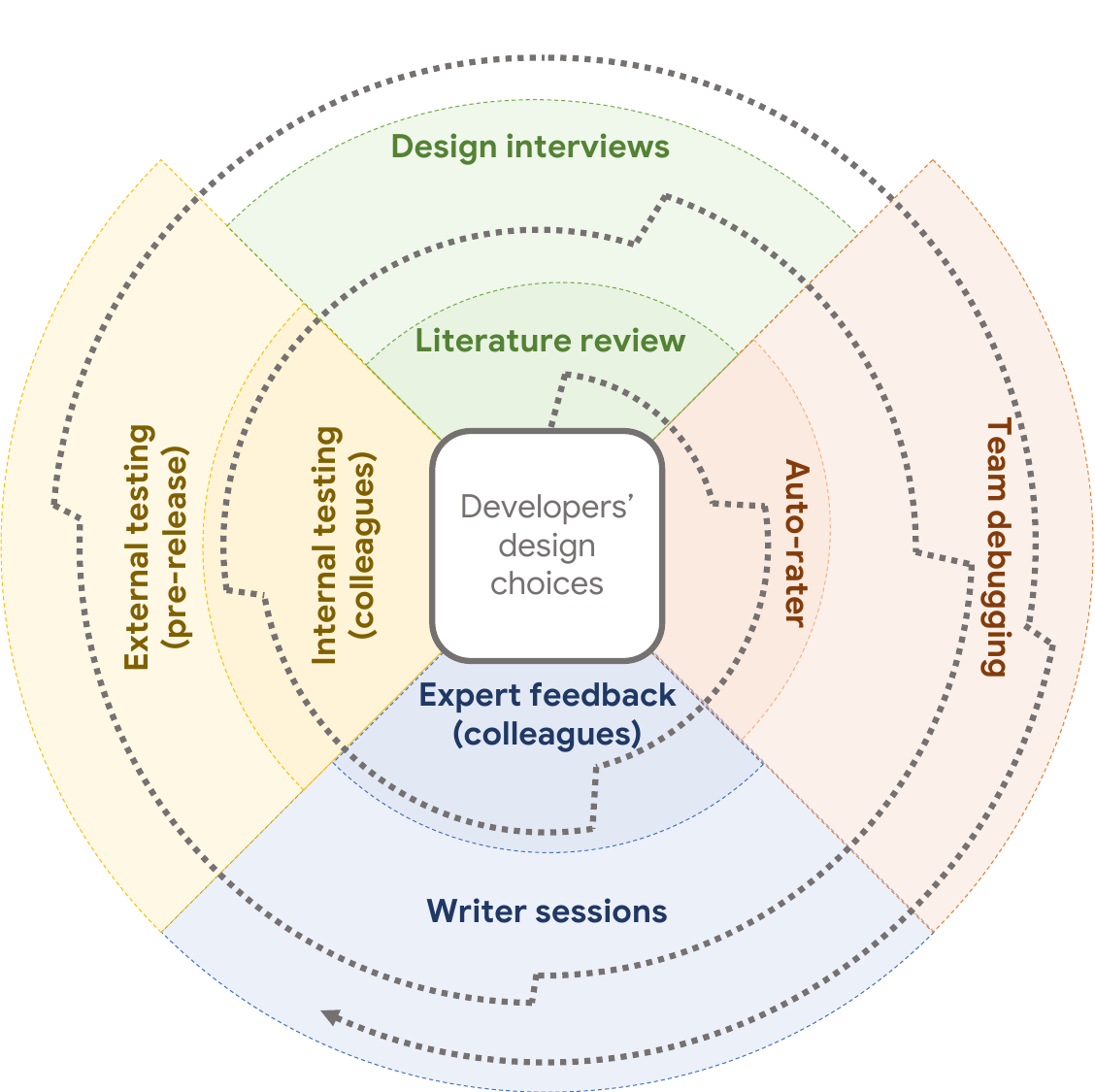}
    \caption{Participatory development cycle.}
    \Description{Schematic showing the development cycle. At the center is a box saying "Developers' design choices". A dashed line departs from the center and turns clock-wise, crossing areas marked "Literature review", "Auto-rater", "Expert feedback (colleagues)", "Internal testing (colleagues)", "Design interviews", "Team debugging", "Writer sessions", "Expert testing (pre-release)" and "Design interviews".}
    \label{fig:dev-cycle}
\end{figure}

\subsubsection{Development process.}
The chronological order in which this development and evaluation cycle happened allowed us to iterate between internal debugging within the wider \dramabox~team and colleagues, external feedback on design choices, large scale internal feedback on the updated system, and the external writing study and external testers (as illustrated by diagram on Figure \ref{fig:dev-cycle}). Table \ref{tab:decisions} lists the design considerations for the UI, informed by literature review (theoretical) as well as the UI design changes made as a results of feedback from interview-based and writer sessions.

\begin{table}[]
    \scriptsize
    \centering
    \begin{tabular}{l l l l}
        \bottomrule
         & Functionality & Design consideration & Source \\
        \toprule
        Theoretical & Regeneration & Continuation, alternative generation & (1) \\
        (literature review) & Rewriting & Instruction-based generation and rewriting & (2) \\
         & Character descriptions & Detailed character descriptions & (3) \\
         & Character motivations & Detailed character motivations & (4) \\
         & Story version control & Undo/redo for story version control & (5) \\
        \bottomrule
         Empirical & Character arcs & Scene-by-scene character arcs & Sec. \ref{subsubsec:characters} \\
         (design interviews) & Script uploading & Uploading, parsing and modifying existing script & (6) Sec. \ref{subsubsec:upload} \\
        \bottomrule
        Empirical & Chat interface & Chatbot interface on the side of the Architect view & Sec. \ref{subsubsec:chatbot} \\
        (writer sessions) & Script critique & Chatbot to provide feedback and critique of script & Sec. \ref{subsubsec:teaching} \\
         & Lock elements & Controls to lock story elements during generation & Sec. \ref{subsubsec:lock-add-remove} \\
         & Add/remove elements & Controls to add/remove story elements in the story plan & Sec. \ref{subsubsec:lock-add-remove} \\
         & Insert/delete scenes & Controls to insert/delete scenes in the story plan & Sec. \ref{subsubsec:lock-add-remove} \\
         & Build-up progressively & Create the story plan in progressive stages (characters, etc.) & Sec. \ref{subsubsec:progressive} \\
         & Scenes/beats count & Suggest the number of scenes/beats from synopsis/plan & Sec. \ref{subsubsec:lock-add-remove} \\
         & Formatting & Formatting specialisation to screenplay or theatre play & Sec. \ref{subsubsec:style} \\
         & Illustrations & Making illustrations optional & Sec. \ref{subsubsec:illustrations} \\
         \hline
    \end{tabular}
    \caption{Theoretical and empirical UI design considerations from literature review, design interview sessions and writer sessions. References: (1) \citep{mirowski2023cowriting,chakrabarty2024creativity,yeh2024ghostwriter,ippolito2022creative}. (2) \citep{yeh2024ghostwriter,ippolito2022creative}. (3) \citep{ippolito2022creative}. (4) \citep{mirowski2023cowriting}. (5) \citep{yeh2024ghostwriter}. (6) \citep{mirowski2023cowriting,ippolito2022creative}.}
    \Description{Table listing the theoretical and empirical UI design considerations from literature review, design interview sessions and writer sessions. Column names are blank, Functionality, Design consideration and Source.}
    \label{tab:decisions}
\end{table}

\subsubsection{Preliminary literature review for designing the \dramabox~UI .}

Before working on the app, we conducted an extensive literature review of previous human-computer interaction and design work on creative writing assistants, in order to identify the key features requested by writers. We prioritized incorporating the possibility of a) rewriting and editing (continuation, elaboration, alternative generations and in-place editing) \citep{mirowski2023cowriting,chakrabarty2024creativity,yeh2024ghostwriter,ippolito2022creative}, and b) instruction-based generation and rewriting \citep{yeh2024ghostwriter,ippolito2022creative}. When working on the Drama Managers, we decided to incorporate c) the given circumstances (who/what/where/why/how) for each character \citep{ippolito2022creative} and d) detailed character motivations \citep{mirowski2023cowriting}. We implemented e) version control of stories to allow the writer to go back to previous versions \citep{yeh2024ghostwriter}. Finally, we take advantage of Gemini's long context to allow the writer to f) feed \dramabox~with their own existing material \citep{mirowski2023cowriting,ippolito2022creative} to build upon or to generate alternative versions.
Potential desirable features that were not included for questions of time or complexity, were: a) side-by-side comparison of written and/or generated material \citep{ippolito2022creative}, b) getting feedback on existing material \citep{mirowski2023cowriting}, c) search and retrieval to understand where some ideas would come from \citep{guo2025pen}, d) constructing a social graph of characters \citep{mirowski2023cowriting}, and e) providing the writer with a graph-based visualization of the stories. Our theoretical (literature review-based) design considerations are in Table \ref{tab:decisions}.

\subsubsection{Targeting the audience for \dramabox.}

Our primary intended audience is \emph{casual} users. Among casual writers, we distinguish \emph{novice} writers (who do not have experience of creative writing and could benefit from learning writing structure) from \emph{amateur} writers (who may have self-published some of their creative writing on a blog, a student journal, a role-playing campaign, but may look for inspiration, world-building and feedback). We also wanted to support \emph{published} writers, in particular TV and cinema screenwriters who are used to working in writers' rooms and are open to using AI tools \citep{mirowski2023cowriting}, as well as theatre playwrights interested in interactive storytelling. We chose to rely on the feedback from expert writers, in particular screenwriters and playwrights, as they are often educators who give classes or write textbooks on storytelling, can articulate the issues with the tool, and thus speak for experts and amateurs alike. Their feedback helped us narrow down the audiences and applications for \dramabox. 

\subsection{Human-Computer Interaction methods}
\label{subsec:hci-methods}

To evaluate \dramabox~with human users, we followed methodologies in previous work on evaluating co-creation and AI writing software, and engaged both \emph{casual} users (i.e. novice or amateur writers) and \emph{professional} writers. We sought feedback from writers among our own colleagues ({\bf internal testers}), recruited external published writers and academics (for 1-hour {\bf design interviews} and 3-hour {\bf writer sessions}) and finally, we publicized the tool to our professional and personal networks of casual and experienced writers ({\bf external testers}). Ethical approvals are detailed in Appendix \ref{subsec:appendix-approvals}.

We designed evaluations of four types: short user surveys (Section \ref{sec:methods-user}), design interviews (Section \ref{sec:methods-design}), longer creativity surveys (Section \ref{sec:methods-creativity}) and mixed-method writer sessions (Section \ref{sec:methods-writers}).

\subsubsection{Short user surveys (internal and external testing).}
\label{sec:methods-user}
\dramabox~users can provide in-app feedback, rating the application using a five star scale, and answering a simplified questionnaire (exact questions are listed in Appendix \ref{subsec:appendix-users}). The survey also collects free-form written user feedback to collect design suggestions from writers and to allow them to comment on the tool usability. The goal of these surveys is to uncover bugs within \dramabox, and to obtain large-scale validation of the tool with a diverse population of external users, including casual and professional writers.

\subsubsection{Design interviews.}
\label{sec:methods-design}
\begin{figure}
    \centering
    \includegraphics[width=1\linewidth]{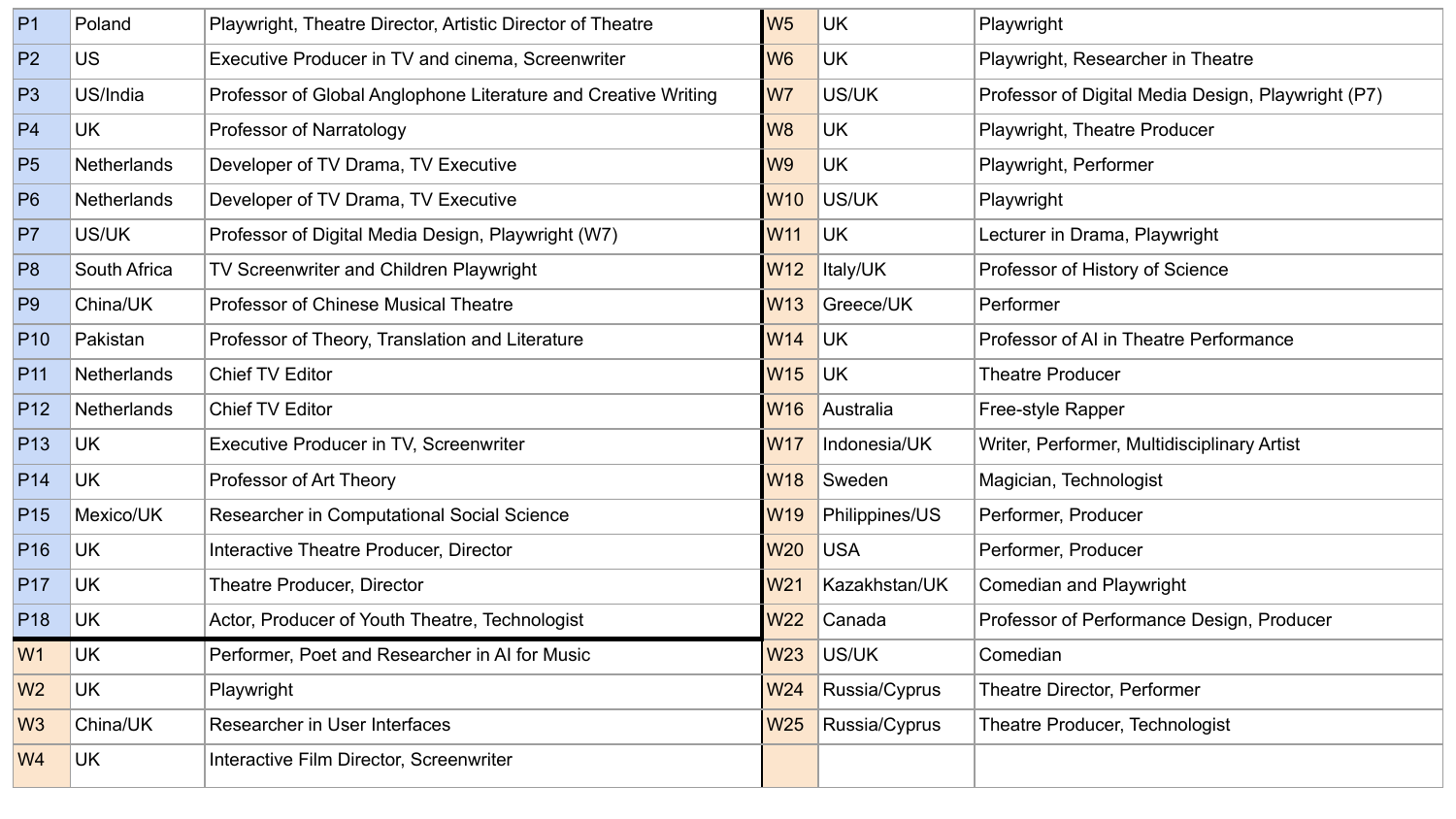}
    \caption{List of 42 participants for the Design Interviews (P1 through P18) and Writers Sessions (W1 through W25).}
    \Description{Table of 43 participants, showing the ID (P1 through P18, W1 through W25), country, and creative writing job description.}
    \label{fig:participants}
\end{figure}

We conduct qualitative verbal feedback recorded during one-hour interview sessions, where we seek feedback on \dramabox~design choices. The 18 external experts (see Fig. \ref{fig:participants}) were recruited using snowball sampling \citep{browne2005snowball,noy2008sampling,naderifar2017snowball}, by reaching to personal networks then leveraging those connections to recruit more participants.

The specific questions asked during design sessions with experts (published writers and academics teaching writing) are listed in Appendix \ref{subsec:appendix-design}. These questions are used to seed the one-hour conversation with guided questions and one to two moderators. Design sessions are recorded and transcribed, then qualitatively coded and discussed by two researchers (who are also authors of the paper).
Inductive thematic analysis \citep{braun2006using,maguire2017doing} (allowing codes and themes to emerge directly from the transcripts) was used to associate feedback and concerns with specific functionalities or assumptions made in \dramabox.

\subsubsection{Creativity surveys (internal experts and writer sessions).}
\label{sec:methods-creativity}
To assess the writer's experience of using \dramabox, to evaluate the creativity support \citep{shneiderman2007creativity} of \dramabox, and to provide detailed feedback on the tool's usability from both casual and published writers, we design an in-depth questionnaire, using questions slightly adapted from~\citet{yuan2022wordcraft,stevenson2022putting,mirowski2023cowriting} and listed in Appendix \ref{subsec:appendix-creativity}.

\subsubsection{Mixed method writer sessions.}
\label{sec:methods-writers}
We conduct mixed methods quantitative and qualitative three-hour writing sessions: the first 90 minutes are spent on presenting \dramabox~and on individual interaction with the tool, then writers answer a creativity support survey, and we conclude with 60 minutes of focus group, similarly to the method used in design interviews (see Section \ref{sec:methods-design}).
Similarly to design interviews, the 25 external writing experts (see Fig. \ref{fig:participants}) were recruited using snowball sampling.
The specific questions asked following writing sessions are listed in Appendix \ref{subsec:appendix-writers}, including questions used to evaluate the Creativity Support Index of the tool \citep{cherry2014quantifying} (based on the NASA Task Load Index \citep{hart1988development}).

%% file: paper_design_quality.tex
\label{sec:quality}
In this section, we analyze the confrontation between the designers' design choices made for algorithmic improvement of stories generated through LLMs and the writers' understanding of story quality.

\subsection{Developer's aim: to create high-quality narratives and illustrations through self-improving generation}

\subsubsection{Hypothesis: can the quality of the story generator self-improve to surprise the reader?}
As designers we responded to the perception that artefacts produced by LLMs can feel ``boring'': despite following instructions correctly, the writing may fail to add enough detail or integrate ideas in a surprising way \citep{bissell2025theoretical,schulz2024narrative}. The need for surprise echoes the long-standing goal of a computationally creative system that can surprise its creator \citep{boden2009computer,boden1998creativity}. One of our aims with \dramabox~was to raise the quality of LLM-generated text by guiding the model through a narrative ``thinking'' process (using ``chain-of-thought'' prompting \citep{wei2022chain}) to write stories that are both consistent and hopefully less formulaic. As a second goal, we also sought to validate our methods and any subsequent updates by providing a repeatable, automated protocol to gauge story quality as we make adjustments to the Drama Managers in \dramabox. The auto-rater (see next section) would be a proxy for a human evaluator. We assumed that we can get a good auto-rater that correlates well with human users by asking the model to evaluate the text with respect to a number of evaluation questions distilled from narrative theory.

\subsection{Self-play and auto-rater developments for \dramabox}
\label{subsec:self-play}

With the auto-rater \citep{zheng2023judging}, we wanted to optimize {\bf narrative quality} (as evaluated by an LLM proxy), {\bf instruction-following} (given the user's requirements) and model {\bf responsiveness} (how quickly it responds). We combined narrative quality and instruction-following into a single {\bf quality} metric, and aimed at expanding the quality-responsiveness Pareto frontier. Our iterative design process was a loop in which \dramabox~generates stories via self-play, then automatically evaluates those stories using the auto-rater, to guide our development of new drama managers.

\begin{figure}
    \centering
    \includegraphics[width=1\linewidth]{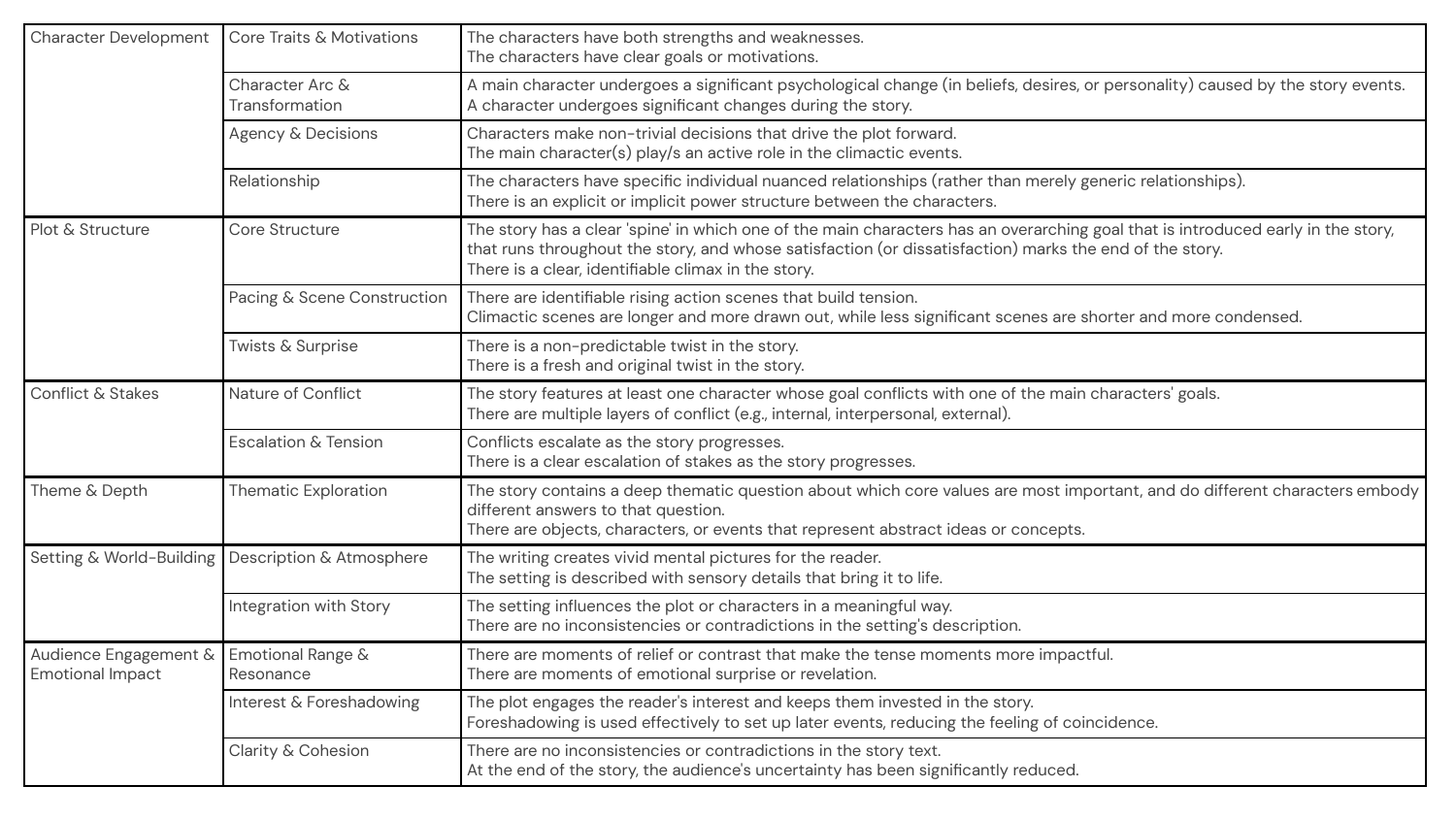}
    \caption{Taxonomy of auto-evaluation questions used by the auto-rater during self-play, and currently reused for writing feedback.}
    \Description{Table showing the taxonomy of auto-evaluation questions. The first column shows main categories, the second column shows subcategories, and the third column shows the question.}
    \label{fig:autorater}
\end{figure}

We implemented diverse self-play agents with different ``personalities'', ranging from hands-off (just asking \dramabox~to keep generating the story) to more proactive agents that edit and revise the story. Self-play agents served the purpose of testing and debugging \dramabox~as well as generating stories for auto-evaluation.

For auto-evaluation, we followed the guideline-based interpretable auto-rater design from \citep{lewenberg2025plan}. Specifically, we constructed a set of 63 guidelines (single-sentence statements distilled from textbooks on the theory and craft of fiction writing \citep{mckee1997story,lowe2000classical,yorke2013into} and from the Torrance Test of Creative Writing \citep{chakrabarty2024art}). Figure \ref{fig:autorater} shows the taxonomy of questions with some examples. The auto-rater prompts an LLM (Gemini 2.0 Flash \citep{team2024gemini} at the time of the app implementation phase) to evaluate each guideline on a 5-point Likert scale and produce a final average score. To mitigate the core problem of LLM-based auto-raters (which are biased towards selecting either the first or last option in multiple-choice questionnaire  \citep{wang2024large}), we enforce the model to generate arguments for each of the ratings from 1 to 5 before generating the final rating.

To validate the effectiveness of our auto-rater, we compared it side-by-side to human annotators\footnote{We followed ethics guidelines of our research institution and employed university-level expert annotators, compensated at a competitive salary.} asked to annotate their preferences between two different \dramabox-generated screenplays, following the design in \citep{huot2024agents}. Given two screenplays, human annotators were asked to choose the better one in terms of plot, creativity, development, language use, and overall quality. We measured the correlation between auto-rater preference and human overall preference\footnote{We use human preference data from \citep{huot2024agents} to compare human preference with auto-rater preference, calculating the Fleiss' Kappa of 0.46 ($p < 0.01$, $N = 150$ participants, $k = 3$ categories), which we as interpret ``to be satisfactory given the subjectivity of our task''.}: the original guideline-based auto-rater scored 0.72 while our improved version scored 0.83.

\subsection{Feedback from writer sessions}

\subsubsection{About the quality of story structure.}
The prominent feedback was that the system excelled at the mechanical aspects of storytelling such as structure, scene breaking and even \nquote{quite abstract and amusing stories}{W11}. W4 \pquote{thought [\dramabox] was terrific [...] in terms of how to break an idea into a perfectly workable scene}. W9 observed that while \dramabox~is \pquote{not great at dialogue; it's really good at setting scenes} and that it \pquote{did better creating an emotional meaningful story arc than it did making it a good play}.

\subsubsection{About the quality and usefulness of LLM writing style.}
Opinions were more contrasted about the stylistic quality of the script. Some writers found \nquote{some great snappy one-liners for a drag queen character [...] I'm stealing that one}{W9}, or praised the \pquote{mainstream} quality \nquote{that lots of people would want to watch}{W4}. A few participants recognized that LLM writing, while of lesser quality, could have a broader appeal which could be valuable and useful: \pquote{it was kind of helping [W4] to really make it into a script of a kind of commercial, regular mainstream sort that lots of people would want to watch}, like \nquote{collaborating with someone who is much more commercially minded}{W2}; \nquote{if a human writer had done that, I would have been pretty happy with that as a first draft}{W4}. 

However, and as W16 acknowledged, \pquote{the majority was actually quite inspiring for me, but I think that's because I top-loaded it so much [with my own content]}. During a lively exchange, writers commented over limitations of \dramabox~(at that time powered with out-of-the-box Gemini 2.5 Flash) in imitating style: \nquote{tried to get it to write something in the style of Brecht [and it failed]}{W6}, \nquote{tried Kronengberg and it failed}{W7}, \nquote{tried Samuel Beckett, which had some success, but actually trying to get it to not write very expositionally, or to actually come up with kind of clownish stage directions, was quite difficult}{W14}, \nquote{tried Pinter and nothing even started (laughter)}{W9}.

Participants found \dramabox~struggled with the organic creativity that were markers of quality writing. The use of traditional LLMs underlying \dramabox~meant that outputs did feel derivative, \nquote{generic}{W2}, \nquote{corny}{W3}, \nquote{creative 101}{W10}. The system's output seemed to reach a stylistic ceiling by producing a mainstream average that writers felt lacked the intentional rule-breaking characteristic of specific literary voices, like \nquote{collaborating with someone who is much more commercially minded}{W2}. This lack of nuance was a limitation for W17, as \pquote{it doesn't understand at all what irony or satire is}.

\subsubsection{About the quality of image illustrations.}
\label{subsubsec:illustrations}
\begin{figure}
    \centering
    \includegraphics[width=0.49\linewidth]{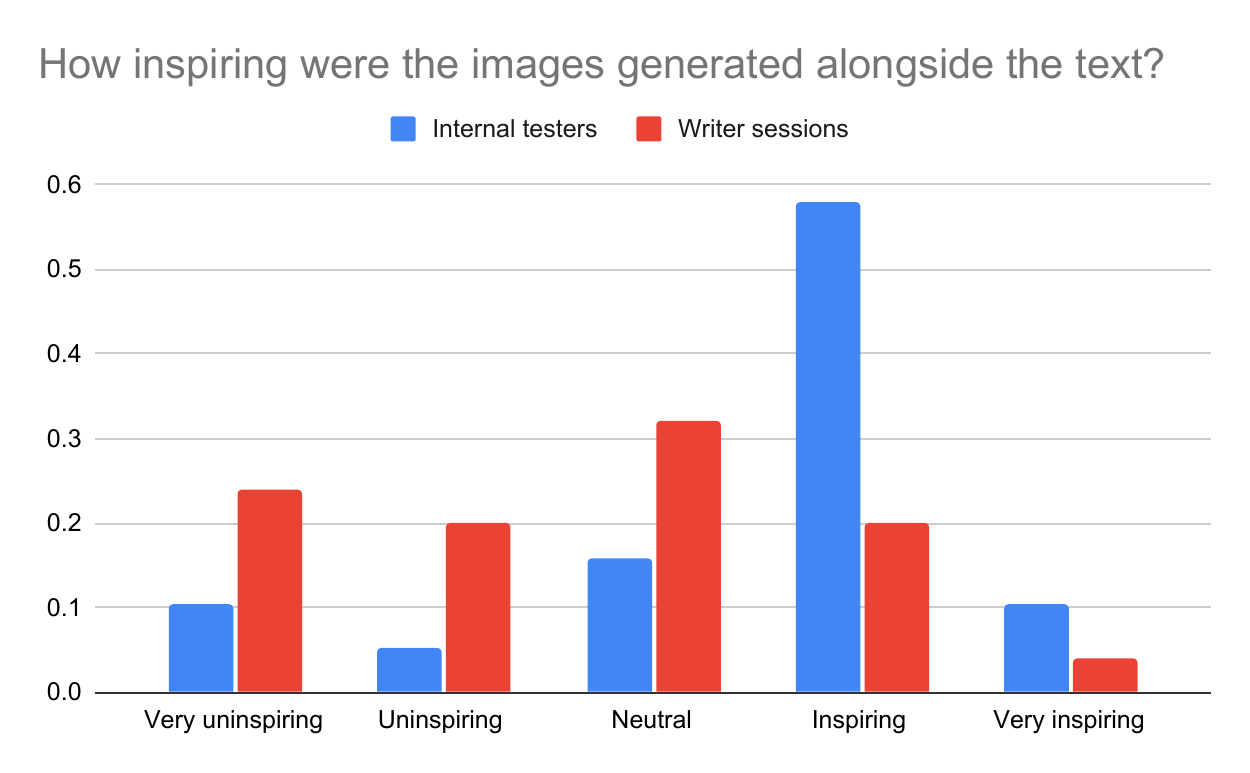}
    \includegraphics[width=0.49\linewidth]{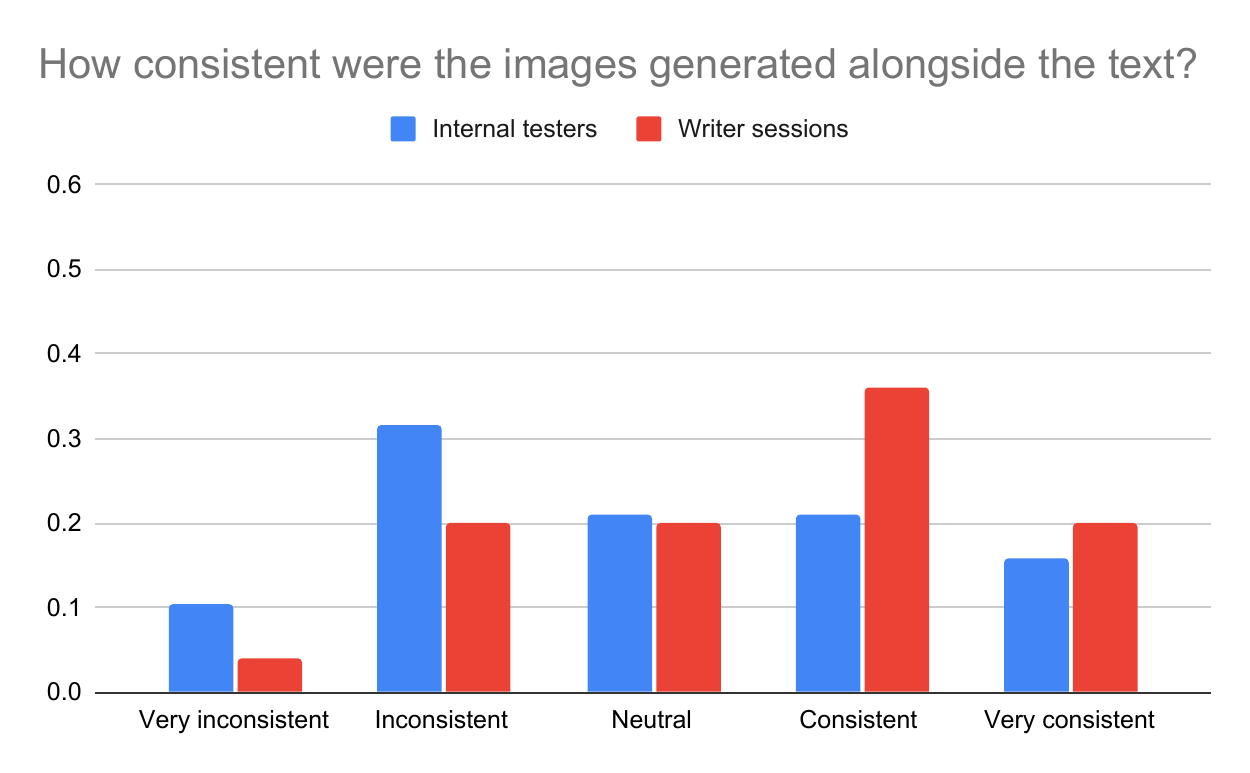}
    \caption{Quality of the illustrations in \dramabox, as a fraction of the responses, for internal testers and during writer sessions.}
    \Description{Two bar chart plots with 5 categories on the X axis and fractions on the Y axis. The titles are "How inspiring were the images generated alongside the text?" (left) and "How consistent were the images generated alongside the text?" (right). Legends show two categories: internal testers and writer sessions.}
    \label{fig:illustrations}
\end{figure}

We report on images even if their quality was not optimized by auto-raters. Whereas internal testers tended to like illustrations generated in \dramabox, writers' opinions were mixed (see Fig. \ref{fig:illustrations}), sometimes questioning their utility for playwriting (W4); image consistency and plausibility were seen as the culprit. As a result, we added a toggle to the app to {\bf make illustrations optional}.

\subsection{Reflection on design choice and iterative development}

There was a tension in the way we (as designers) defined story quality and ``improvement'', and the writers' understanding of it. The system's goal of consistency was sometimes at odds with one element of quality that some writers expected: de-familiarization as a literary device, or the presence of the unexpected.

\subsubsection{Preliminary feedback from design interviews emphasizes desire for the unexpected.}
P2 noted that quality meant \pquote{piercing the predictable version [...] looking for are those branches where you take me into an unexpected direction}, while P1 and P7 explicitly asked for an \pquote{absurdity dial} or more \pquote{non sequiturs}. An auto-rater like ours, designed to reduce ``inconsistencies'', might inadvertently filter out the glitches, productive frictions or absurdities that experienced writers use as creative springboards. Arguably, this design decision supports novice writers.

\subsubsection{Iterative development of feedback on script that respects the author's originality.}
As evidenced in Section \ref{sec:creativity}, participants in the writing sessions wanted feedback on existing scripts and story structures. We therefore re-used the auto-rater (Fig. \ref{fig:autorater}) to build a new feature in \dramabox~to critique and propose improvements to work-in-progress script. The chatbot formulates them as mere suggestions, to respect the author's originality and intent.

%% file: paper_design_architect.tex
\label{sec:gardeners-architects}

\subsection{Developers' hypothesis: depart from chat-based tools to support \emph{Gardeners} and \emph{Architects}}

\subsubsection{Developing a writing app that allows to switch between Gardener and Architect workflows.}

While dominant AI writing tools rely primarily on a dialogic chatbot interface, we aimed to move toward a more immersive environment that supports the two primary writing archetypes popularized \citep{flood2011getting} by George R R Martin\footnote{\nquote{The {\bf architects} plan everything ahead of time, like an architect building a house. They know how many rooms are going to be in the house, what kind of roof they're going to have, where the wires are going to run, what kind of plumbing there's going to be. They have the whole thing designed and blueprinted out before they even nail the first board up. The {\bf gardeners} dig a hole, drop in a seed and water it. They kind of know what seed it is, they know if planted (sic) a fantasy seed or mystery seed or whatever. But as the plant comes up and they water it, they don't know how many branches it's going to have, they find out as it grows. And I'm much more a gardener than an architect.}{George RR Martin}}: \emph{Gardeners}, who write through improvisational discovery, and \emph{Architects}, who build from a structured blueprint. Our goal was to create a system that maintains high-level narrative structure while allowing for granular, moment-by-moment creative immersion.

To support Gardeners, \dramabox~allows users to generate narratives one beat \citep{mckee1997story} at a time. We hypothesized that this beat-by-beat generation situates the Gardener writer within the immediate flow of the scene. For Architects, we implemented a \emph{Story Plan} view to visualize the entire narrative arc. This view facilitates a hierarchical search through the story space, allowing writers to modify high-level plot points that then propagate down to the script level. 

In addition to manual edits directly in the text, the writer can conversationally edit the story plan and the scene plans, allowing for rapid iteration. User intent can also percolate upstream: if a user writes an action in the script in a scene that contradicts the larger story plan, the app automatically adjusts the scene and story plans. These automatic updates to the story plan can be disabled, so that small changes to an individual scene do not affect the overall story structure.

By providing this hierarchical, multi-level editing environment, we hypothesized that we could offer writers a more precise dramaturgical steering wheel. By  combining the Gardener and Architect interaction modalities we had hoped to provided different levels of agency for different writer styles.

\begin{figure}
    \centering
    \includegraphics[width=0.49\linewidth]{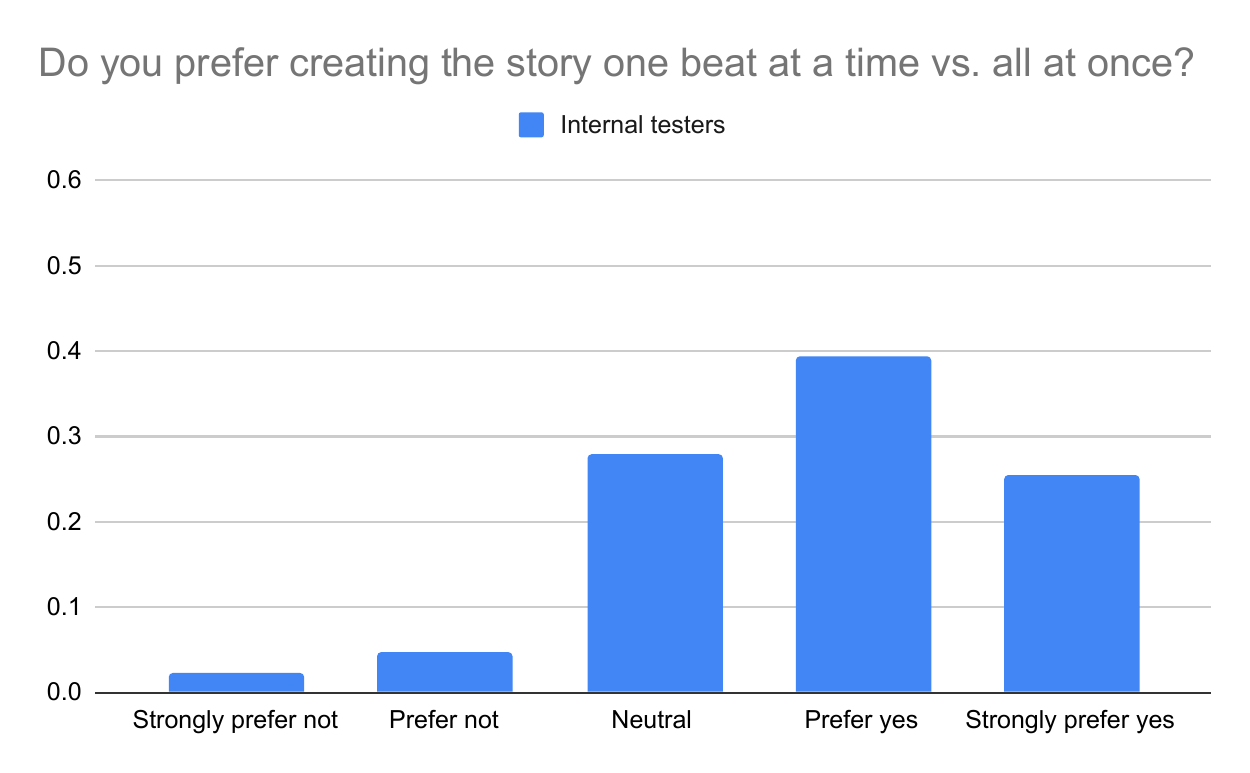}
    \includegraphics[width=0.49\linewidth]{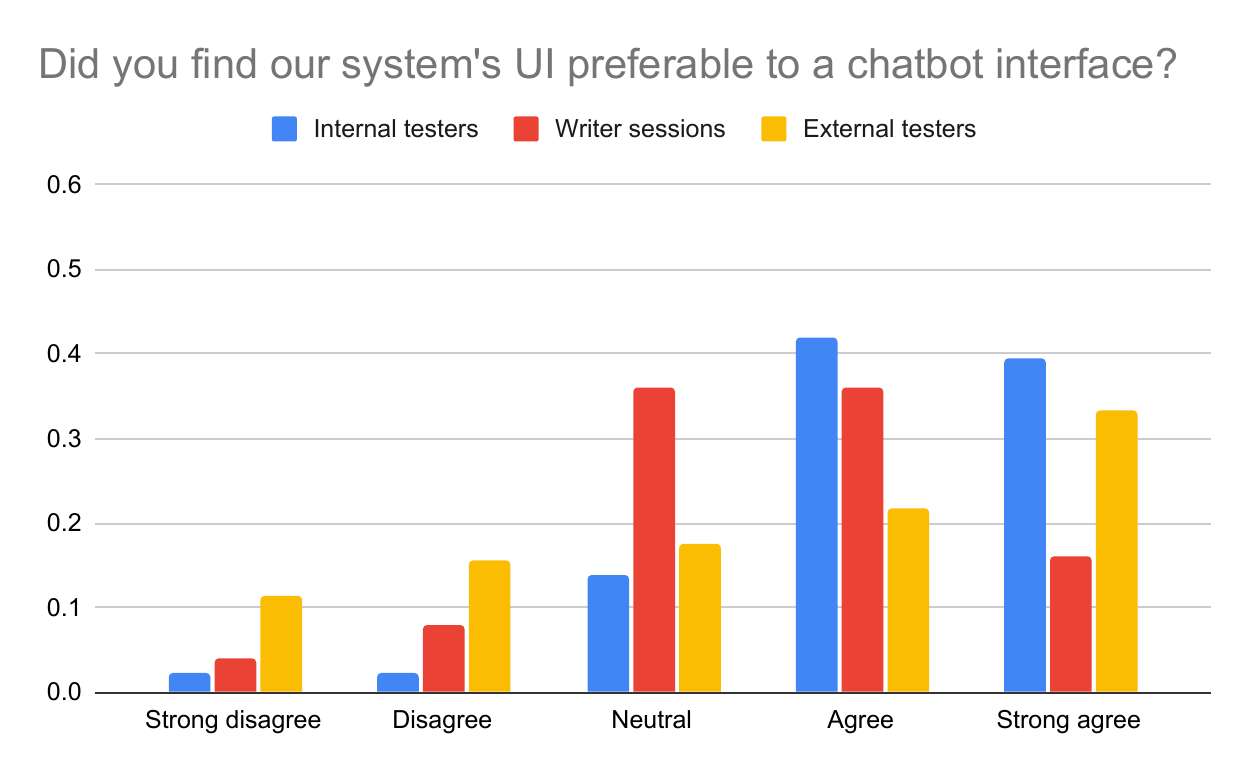}
    \caption{Preferences for the UI, as a fraction of the responses, for internal testers, external testers and during writer sessions. Left}
    \Description{Two bar chart plots with 5 categories on the X axis and fractions on the Y axis. The titles are "Do you prefer creating the story one beat at a time vs. all at once?" (left) and "Did you find the \dramabox~UI preferable to a chatbot interface?" (right). Legends show three categories: internal testers, external testers and writer sessions.}
    \label{fig:chatbot}
\end{figure}

\subsection{Feedback from design interviews and writers sessions}

\subsubsection{About the pertinence of the ``Architect'' and ``Gardener'' concepts.}
Experts in narratology initially validated these metaphors (P2, P4, P5, P6, P7): \nquote{Everyone says there's two tribes of writers: there's the plotters and the pantsers. So, the plotters have everything really carefully worked out and the pantsers fly by the seats of their pants. The hierarchical approach helps you get that kind of baseline structure from the get-go [...] It's becoming a scaffold for other scaffolding [...] You can then build other, sort of mezzanine levels that are also critical.}{P4}.

Experts noticed how the Architect view enabled to \nquote{restructure all of the narrative to make sense}{P7} and to re-plan stories \nquote{with a click of a button}{P8}.  W11 said: \pquote{I teach structure [...] and the Architect mode did allow me to do that} and W7 remarked that the \pquote{ability to track beats over a storyline indirectly led to new ideas about how to structure a story}. W15 likened the Architect mode to \pquote{creating your character sheet in Dungeons and Dragons}.

Conversely, some writers made parallels between the Gardener UI and a \nquote{computer game}{W18}. W4 found the gardener mode \pquote{fun because it's sort of more granular [...] moment by moment kind of co-creation} that \pquote{actually came up with things that I found more surprising and interesting}.

\subsubsection{About the rigidity of the story plan structure in Architect mode.}
\label{subsubsec:lock-add-remove}
The actual experience of using the tool revealed a significant friction between systemic rigidity of our modes and creative fluidity. While writers lauded the scaffold provided by the hierarchical view, they simultaneously found the interface \nquote{rigid}{W12}. Some writers (W1) commented on struggles with steering the story using the Architect mode. W12 felt the Architect mode] was \pquote{very rigid in those three columns}. W13 noticed that \pquote{a mere addition of a character without messing up the rest of the plot doesn't seem to be feasible} as \nquote{every time I have had it rewrite a thing, it would rewrite the whole structure}{W7}. W12 wanted to be \pquote{able to be more operative on changing bits one by one}, to \pquote{add another character}, or \nquote{add goals or world building elements}{W22}, and give \nquote{suggestions of number of beats}{W2} as when \nquote{I am starting a story, I do not know how many scenes I want}{W23}.

As a consequence of that rigidity, W5 noticed that \pquote{it wasn't always perfect at understanding cause and effect; so, if I changed something in scene 3 a couple times, it would still have a scene in scene 6, where that character hadn't been locked up as they were in scene 3} and W23 remarked that \pquote{when you refresh a story, it does not refresh the picture, so the picture no longer matches the story.} W7 and W14 wanted more clarity when they make bottom-up edits at the script level.

Writers made specific suggestions for improvements as \nquote{it can be useful actually if there are more controls}{W17}.
Requests included more progressive generation of the story into scenes and beats (W5), \pquote{not having five audience response questions that I feel I need to input [to avoid] weird scaffoldings coming in}, \nquote{to be able to add a scene in between}{W18} or to \nquote{add this extra character in the existing plot}{W13}, an \nquote{'inspire me' button that would generate a goal idea based on the context of all the other goals}{W16}, to \nquote{annotate the final script}{W18}, \nquote{a scratch pad to copy paste ideas}{W23}. Therefore, we {\bf added controls to easily lock, add and remove elements} of the story plan.

\subsubsection{Visual complexity of Architect mode vs. existing writing tools and workflows.}
\label{subsubsec:progressive}
\dramabox~presented writers with unfamiliar visual complexity: \nquote{I thought it was too complex. I wanted a more simple breakdown [...] I didn't want so much detail}{W6}; \nquote{as an artist without IT background, [I found it] very stressful to use this tool}{W24}. W1 noticed \pquote{it wouldn't work on a mobile [...] you've got quite a complicated UI}. For some professional writers, the cognitive overhead required to maintain our three-column architecture could disrupt the flow the tool was meant to support. 

Part of the writers' confusion was due to {\bf missing tutorials explaining some of the functionalities of the tool}, which we have added. Some writers also recommended to reduce unfamiliar visual complexity and to adopt the UI and layout of industry standard writing software like \emph{Final Draft}, \emph{Highland} or \emph{Scrivener} (W2, W5, W12, W22, W23). W12 and W23 suggested white on black color schemes: \pquote{Writers want light colours, dark is very tech; see what Scrivener or Goodreads do}{W23}. We therefore simplified the Architect mode by {\bf building the story plan elements progressively} (e.g., first characters, then world-building questions, etc.)

\subsubsection{About modularity vs. linearity and top-down structure in \dramabox.}
\label{subsubsec:bottom-up}
The Drama Managers treat a story as a unified object, whereas writers treat it as a collection of modular, loosely coupled experiments. This meant that writers (even architects) did not begin with a rigid plan to the level of granular detail we had set out: \nquote{I am starting a story, I do not know how many scenes I want}{W23}, or \nquote{I don't always write linearly}{W2}, and W13 noticed \pquote{it was easier in terms of modification to go top-down than bottom-up [...] It would be nice [...] to manipulate the lower level and see how this can affect the structure at the high level}. Most writers wanted more flexibility and \nquote{more controls}{W17}, and to see their experiments locally --- like a \nquote{scratch pad}{W23} --- without cascading entire changes throughout the story plan. The systemic coherence which we had intentionally designed, created a loss of agency (W1, W2, W5, W7, W13, W14, W18) as it was impossible to do \nquote{a mere addition of a character without messing up the rest of the plot}{W13}. {\bf We will improve bottom-up workflow in future work}. 

\subsubsection{Comparison with standard chat interfaces.}
\label{subsubsec:chatbot}
In general, writers and testers found the \dramabox~UI --- \nquote{set up specifically for writing scripts}{W4}, \nquote{easier [to] just jump back and forth}{W3} and \nquote{more descriptive than on Gemini}{W20} --- preferable to a chatbot interface (Fig. \ref{fig:chatbot}). But despite our initial aim to move past the chatbot, some writers regretted the \nquote{fluidity}{W5} of a dialogic interface, sometimes out of usage (W10, W21, W23): \nquote{I already know how to think in a chat app, so I found it easier}{W23} and W21 said: \pquote{with Gemini what I do a lot is make it list things}. Writers sometimes missed the messiness and non-linearity that our structured UI inadvertently suppressed.

W22 suggested an improvement to \dramabox~(\pquote{if there was almost like a chatbot alongside it, or some kind of ability to jump between workflows, then it would fit more into existing writing workflows}) and W18 found a chatbot would be \pquote{complementary}. We therefore {\bf added a chat interface} to \dramabox.

\subsection{Iterative development and follow-up validation}
As a consequence of the initial writer sessions, we implemented multiple changes to the UI (see Table \ref{tab:decisions}).
\begin{itemize}
  \item A {\bf ``lock'' mechanism} that allows writers to manually anchor specific narrative elements (protecting them from global changes and giving the story a modular scaffold), {\bf ``add''/``insert'' and ``delete'' buttons} for lists of elements and for scenes.
  \item We modified the Architect mode to {\bf build story elements progressively}, to reduce the reported \pquote{visual stress} of complex UI. Users can now define characters etc. sequentially before the full three-column view is revealed.
  \item We re-integrated a \nquote{complementary}{W18, W22} and collapsible {\bf chat interface} alongside the story plan.
\end{itemize}

In follow-up sessions, writers commented positively on the flexibility introduced by these UI changes (W4, P18).

%% file: paper_design_hierarchical.tex
\label{sec:hierarchy}

In this section, we investigate if we can rely on our hierarchical Scene-Beat story plans to structure stories according to culturally diverse storytelling traditions and for various domains from screenplays and theatre plays to novels.

\subsection{Developer goal: to define a story structure as flexible and culturally relevant as possible}

\subsubsection{Selecting narrative elements that accommodate a broad range of storytelling traditions.}

As reviewed in Section \ref{subsec:storytelling}, narrative experts \citep{mckee1997story,yorke2013into} often recommend hierarchical plans in which stories are divided into scenes, and scenes are divided into beats. We therefore implemented a hierarchical Scene-Beat model, whose advantage is to force the LLM to generate a structured plan first (ensuring global narrative coherence), but which also introduce Western biases.

In order to make \dramabox~usable for various narrative traditions, we decided not to enforce specific story grammars in the plot, and focused instead on character arcs, suspense and surprise, and audience expectations (Section \ref{subsec:drama-managers}).

\subsubsection{Aiming towards culturally-situated narratives is a challenge for LLM-based systems.}
Writers observed that LLMs tend to reinforce certain cultural values in a way that erases the writers' individual cultural identity. For instance, instruction-tuned LLMs were shown to sometimes censor minority speech because of misspecified safety concerns \citep{mirowski2024robot} (therefore reinforcing hegemonic viewpoints) or to homogenize towards Western styles \citep{agarwal2025suggestions} (removing cultural nuance).

To avoid reinforcing these biases, we do not fine-tune the LLM and do not include Western examples in prompts.

\subsection{Feedback from design interviews and writers sessions}

\subsubsection{Praise for the possibility to work on characters.}
\label{subsubsec:characters}
Some participants praised how Drama Managers were \nquote{character-led}{P4}, asked the right questions about character development (P3, P7, P9) and allowed to \nquote{go opposite to audience expectations}{P9}.

\subsubsection{Participants' structural concerns about resorting to traditional narratology theory.}
Several participants (like P5, P6, P13 or P16) found not all \dramabox~questions to be relevant in their own practice. In particular, playwright P1 criticised our choice of predicting what the audience feels or learns, as \pquote{it makes \dramabox~the opposite of the ``Open Form'' [...] that allows for as many interpretations as possible}. Similarly, \nquote{meeting anticipated audience goals [...] does not align with with contemporary playwrighting}{P1}.

\subsubsection{Writers observed that \dramabox~is stylistically biased to screenwriting.}
\label{subsubsec:style}
Writers noticed that \dramabox~had a \nquote{screen-writingness}{W12} and was \nquote{very filmic in the way that it's written}{W11}, to the point where W11 and W6 \pquote{tried to get it to do it in a playwrighting format, it just kept going back to screenplay}. Writers wanted to be \nquote{able to be really specific about what medium [they] are trying to create}{W5} and W4 recommended \pquote{a simple drop down saying this is a stage play, this is a film}. We {\bf added stylistic and formatting specialisation to screenplay, theatre play and novel}.

\subsubsection{Participants warned that \dramabox~risks reproducing only well-known Western storytelling structures.}
Participants warned us about underlying Western ideals of the narrative structures we relied upon, which then meant that non-Western forms of storytelling could become mere embellishments on top of a Western structural spine, or that there could be gaps in Chinese literature (P9). In explaining the differences in narratological structures, P3 said that \pquote{Aristotle's Poetics lays a lot of stress and emphasis on causality [...] whereas Nāṭya Shāstra thinks about plot from the perspective of affect much more strongly [...] In the more Euroamerican cultures [...] there is a lot more emphasis placed on the interiority of characters [...] whereas in the South Asian context I would say there is a lot more emphasis on events.} P10 concurred, adding \pquote{I usually say <<Arabian Nights>> because that's the closest we can get to what it means to be in the Dastan world for a Western anglophone audience.} In particular, participants argued that the training data, \nquote{these tools, necessarily grounded in the past}{P1, P4, P16} and \nquote{implicitly rooted in Campbell or Aristotle, systematically missed out feminist, minoritarian, and queer narratological approaches}{P4}, or \nquote{mythemes}{P14} from marginalized cultures.

\subsubsection{Writers observed that LLMs still reproduce cultural stereotypes.}
\label{subsubsec:stereotypes}
During expert interviews, P3 warned us that writing styles are culture-dependent: \pquote{If I'm writing as a South Asian origin writer trained in a couple of South Asian languages [...] am I imagining a South Asian audience that reads English, versus an American or European audience reading English? [...] There is a kind of expectation of a little minimalism of language in the American market versus the South Asian market. [...] There are different ways of storytelling even within the Euro American context.} P8 emphasized that cultural specificity is complex to define and to capture. The responsibility is the writer when writing culturally-specific text: \nquote{I'm very hesitant to write this play because when I write the lingo, it can either work or it can come across as offensive. And that to me—especially because of our history in the country—there is a bit of sensitivity about how we engage or how we portray people from various cultures}{P8}. One cannot \nquote{just translate Dastan-e-Amir Hamza into English}{P10}; even evaluation should be done by \nquote{someone from the culture itself}{P15}.

As we expected, we observed during writing sessions that the underlying LLMs added to \nquote{stereotypical}{W24} Western biases during script generation. For example, the LLM could default to hegemonic social groups: \nquote{people were white and middle class when I deliberately pulled back on not giving any kind of indicators}{W2}, \nquote{Caucasian western culture names}{W3}, \nquote{first choice always defaulted to white and cisgendered}{W6}: \nquote{I very intentionally was trying to not describe the person, and even when I established who it was using gender-neutral terms, it really wanted it to be [...] a 40 something white male}{W22}. In the LLM outputs, \nquote{character descriptions of female characters were very stereotypical like timid, quiet, shy, loving housemaker}{W5} or \nquote{gentle}{W20}, vs. \nquote{high-status}{W11} male: \pquote{I was trying to make the mother character evil [...] but it refused to make it evil}{W20}. Subtle stereotypes manifested across intersectional dimensions, which W21 ascribed to a fear \pquote{of offending}, e.g., \pquote{I was trying to make her kind of sassy, like a lot of central Asian older women that I know in my life, and it kept trying to make it very bland [...] come on, we can make her more sassy!} or \pquote{<<Can we make it a Muslim character?>>, and what it did was give it a generic Muslim name, but then nothing else was added to the story}.

Debiasing LLMs is an ongoing field of work. A few writers noted some successes, e.g., \nquote{a lovely narrative about Pinocchio coming out as non-binary and maybe trans through a metaphor about the lines in your wood being different}{W9}. W6 \pquote{hoped that eventually it would start to show you [diverse characters], even if [...] we aren't enlightened enough at the moment}.

\subsection{Further development}

Acknowledging the problem of cultural flexibility needed in storytelling, we tried to design the Drama Managers to be sufficiently general to support non-Western forms of storytelling, taking only narratology elements that were common to most cultures \citep{lowe2000classical}. Further adaptation will be needed, with {\bf bespoke Drama Managers} that impose even fewer constraints on character, audience or narrator affective goals and perceptions, to express more cultural nuances.

%% file: paper_design_creativity.tex
\label{sec:creativity}

\subsection{Developers' hypothesis: convergent iteration could be used in the writing process}
\label{subsec:convergent-iteration}

\subsubsection{Search, finding, discovery and remixing metaphors.}
Our fourth design choice was rooted in the philosophical shift from ``creating'' a story to ``finding'' one. We drew inspiration from Jorge Luis Borges' \emph{Library of Babel} \citep{borges1944ficciones,borges1998library} containing all possible books of a certain alphabet and number of pages. We framed the writing process as a form of convergent iteration \citep{crawford2004chris} in the story space defined by a high-level story plan, in-depth scene plans and story script.

We hypothesized that by providing persistent, editable artifacts at multiple levels we could support a ``remixing'' paradigm, allowing the user to refine and clarify their ideas over time. According to art critic Lev Manovich, modern media and datasets have enabled \emph{remixability} and encouraged remixing \citep{manovich2005remixing}, a major creative paradigm when using generative AI. \citet{qadri2025ai} documented the process of visual artists iteratively training, generating and \pquote{hacking}
successive generative AI models, while \citet{buschek2024collage} used the metaphor of \pquote{collage} in AI-assisted writing.

We thus positioned the writer as a ``discoverer'' of possible narratives among the shared myths and folk psychologies embedded in LLM training data. We assumed that searching in story space could be useful to both novice and experienced writers. Indeed, the long memory context in underlying LLMs (2M word tokens in Gemini 2.5 \citep{comanici2025gemini}) allows users to input excerpts of their own writing, and to combine their own words with the knowledge base implicitly stored in LLMs, linking AI-assisted writing with so-called AI-driven \emph{database art} \citep{manovich2024make}. 

\subsubsection{Anticipated concerns due to the convergent iteration hypothesis.}

These AI- or remixing-based workflows are however very particular to a small number of writers, and our main questions were how writers would conciliate a top-down hierarchical generative AI tool with their own craft.

First, we hypothesized that writers may feel a lack of human agency and see the final output as ``AI-generated'' or ``AI-ghostwritten'' \citep{draxler2024ai,yeh2024ghostwriter}, questioning their ownership when using such a tool.
Second, we feared that the search and discovery process in AI-generated content might deprive the writers from creative decision making, as suggested by \citet{kreminski2024dearth}.
Third, we wondered if writers will mention risks of cognitive deskilling due to AI usage \citep{kosmyna2025your,kumar2025human}, discussed beyond academic circles at the time of the study.

Desire for creative control and human agency have been documented as a key requirement for creative writing assistants \citep{guo2025pen}, and we anticipated that different writers might want to limit AI assistance to specific functions, such as \emph{guided exploration}, \emph{active co-writing} or \emph{critical feedback}, potentially preferring assistance during the \emph{planning} phase as opposed to \emph{translating} their ideas on paper \citep{reza2025co}. Our large scale study confirmed these previously-documented concerns.

\subsection{Feedback from design interviews and writers sessions}

\subsubsection{Quantitative results suggest the convergent iteration hypothesis works best for novice writers or screenwriters.}
\begin{figure}
    \centering
    \includegraphics[width=0.49\linewidth]{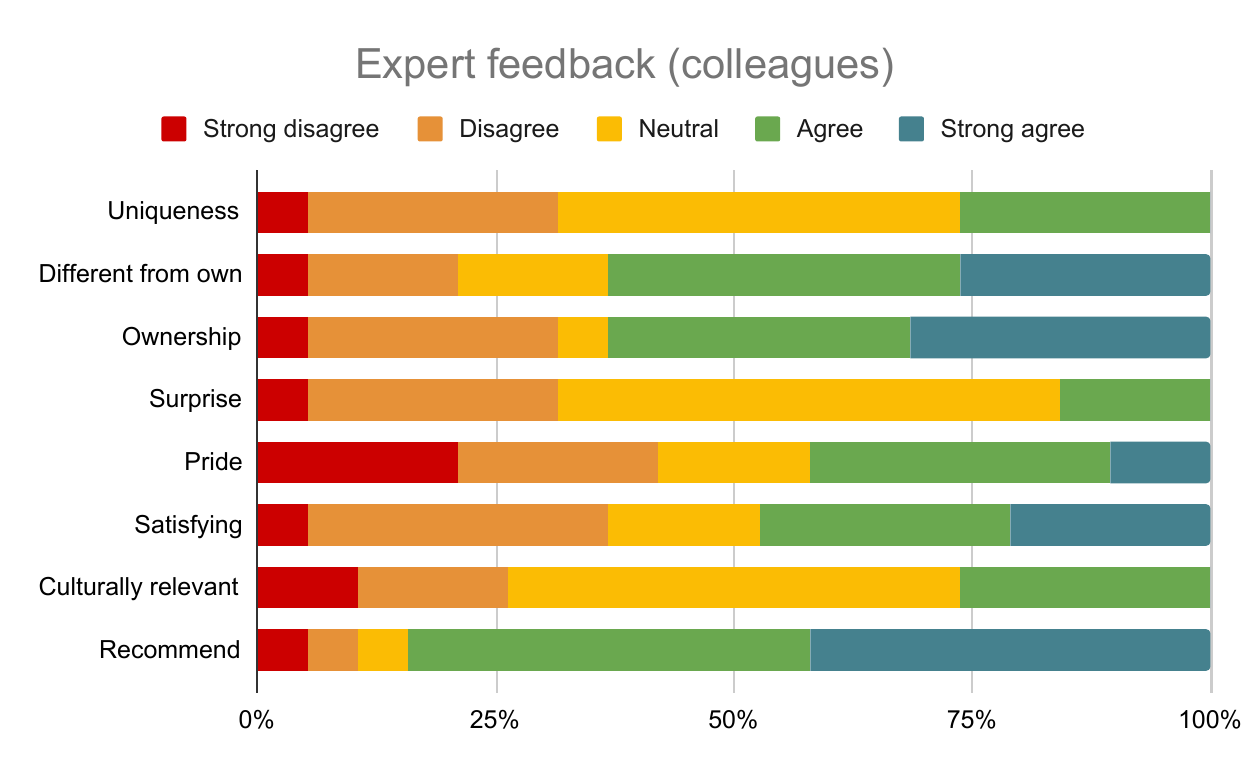}
    \includegraphics[width=0.49\linewidth]{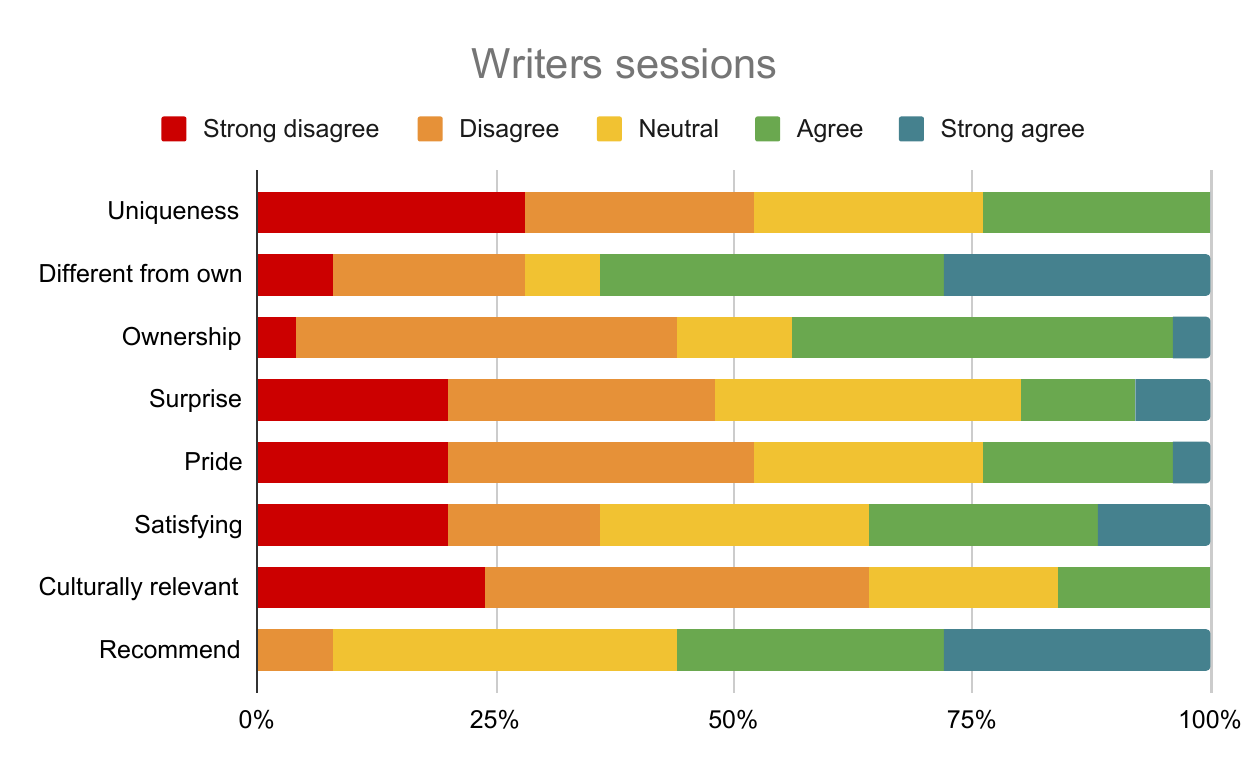}
    \caption{Participants' responses to creativity surveys on a color-coded Likert scale from strongly disagree (red) to strongly agree (blue). Left: responses by 19 colleagues with writer experience. Right: responses from 25 participants in the writers sessions.}
    \Description{Two cumulative horizontal bar charts showing Likert responses to 5 color-coded categories. The Y axis shows "Uniqueness", "Different from own", "Ownership", "Surprise", "Pride", "Satisfying", "Culturally relevant", "Recomment". The titles are "expert feedback (colleagues)" (left) and "writers sessions" (right).}
    \label{fig:likert}
\end{figure}

On the quantitative creativity survey summary on Figure \ref{fig:likert}, we see that opinions of experienced writers about \dramabox's support of the creative process (feeling of surprise,  ownership, pride, satisfaction, uniqueness or cultural relevance of the generated text) were mixed, suggesting the tool would satisfy only a subset of writers (or that it would need improvements). Figure \ref{fig:csi} shows that the Creativity Support Index computed among the pool of 25 creatives (CSI, defined in Appendix \ref{subsec:appendix-creativity}), broken down by the writers' self-declared domain of expertise, was the highest for participants who declared them as ``novices'' or for those who had worked in the film industry (and the lowest for theatre makers).

These quantitative results suggest that the convergent iteration in \dramabox~could be useful for casual writers or as part of the process for screenwriters' workflow (see discussion Section \ref{ref:subsec:opportunities}), as we elaborate upon in the next section.

\begin{figure}
    \centering
    \includegraphics[width=0.49\linewidth]{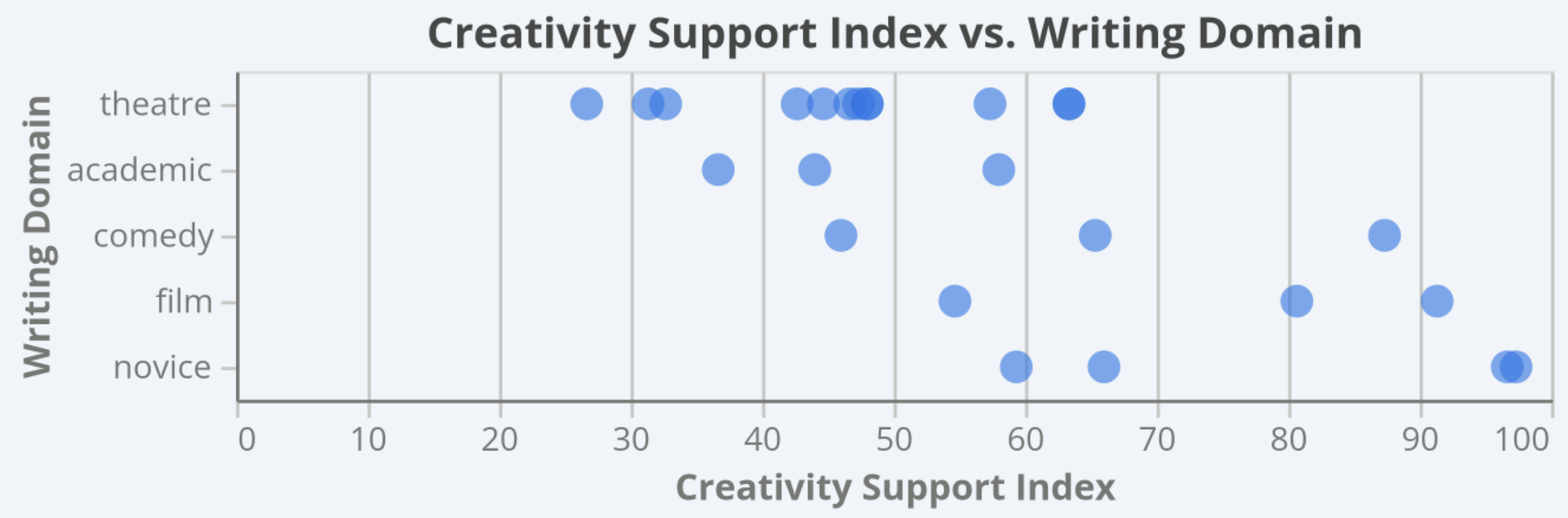}
    \includegraphics[width=0.49\linewidth]{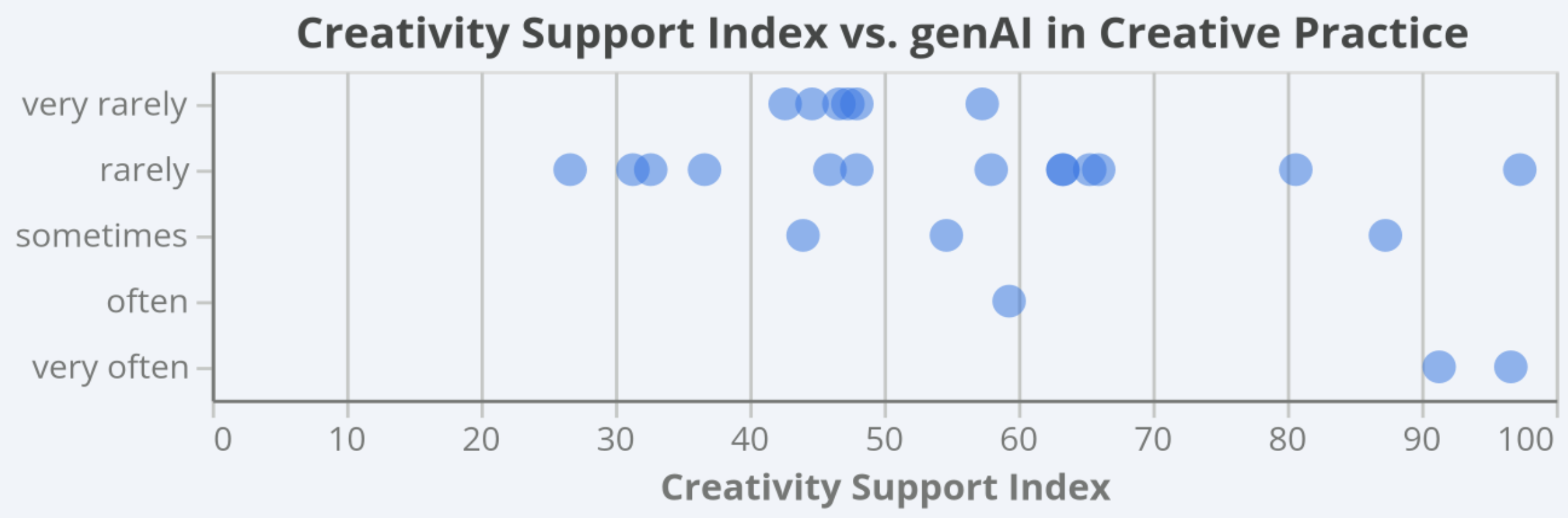}
    \caption{Creativity Support Index \citep{cherry2014quantifying} (defined in Appendix \ref{subsec:appendix-creativity}) of \dramabox, clustered by writers' self-declared domain of expertise (left) and experience of genAI in creative domains (right).}
    \Description{Two scatter plots with the Creativity Support Index in X axis and Writing domain on Y axis (left plot) or GenAI in Creative Practice (right plot).}
    \label{fig:csi}
\end{figure}

\subsubsection{Utility in the ideation process, from inspiration to world-building.}
Writers validated that a tool like this which allows for convergent iteration can be useful towards producing or getting past \nquote{a bad first draft}{W2, W4, W22}, for \nquote{initiating ideas}{P14, P15}, \nquote{background notes for a writer}{W2}, \nquote{helpful for planning the beginning of story}{W23} or \nquote{encouraging to write}{W3}.

It helped writers with \nquote{good world building}{W15} and inspiration for different ways to formulating (W18) or shifting (W7) a story idea, as a sort of \nquote{sounding board}{P2}, that \nquote{give [them] more things that could happen}{W2}.
Some writers imagined using this tool for narrative world-building or ``playground'',  \nquote{an AI, in which you can create characters and a setting, and then you just let those characters loose in that AI world, and see what on the page creates conflict. How do these characters talk?}{P5}. \nquote{I was giggling a lot because some of it was so bizarre that it really appealed to me}{W11}.

The tool then functioned more at the level of \nquote{process}{W5} than at the level of outputs.  \nquote{If I were using it, it would be at the very beginning [...] ideas to bounce off and then I would never use it again in production}{W9}, and \dramabox~would enable \nquote{iterative working out early initial ideas, and then you can depart from the system}{W14}. Copy-pasting dialogue generated by LLMs (\nquote{the easiest part of the equation}{P2}), without completely rewriting it, was seen as \nquote{the most pessimistic}{P2} way of using AI (P1):
\nquote{my intuition would be to then take that first draft, put it into a normal script editor and then change it myself in terms of the dialogue}{P8}.

Our takeaway from the discussions with professional writers was that this tool was perhaps more useful as a point in the creative process, such as \nquote{background notes}{W1}, \nquote{enhancement to an existing script}{W2}, \nquote{analysis of the structure}{W7, W25} of an existing piece of text, help with planning (W5). Writers wanted this tool \nquote{to help me write as opposed to writing it in my place}{W22}.

\subsubsection{Critique of the need to solve the ``blank page'' problem with technology.}

Several writers questioned if the \emph{blank page} is a problem to be solved through technology: to seek inspiration, \nquote{leave your phone at home and go for a walk. [...] Let your mind be bored}{P13}. Some writers found the cognitive load of convergent iteration very high: \nquote{Reviewing these generated versions is more time-consuming than writing a scene from scratch [because] the program generates a high volume of amateur-level material that lacks self-criticism}{P1}.

\subsubsection{Writers' reflections on LLMs and authorship during convergent iteration.}
\label{ref:subsec:authorship}

While probing our vision of convergent iteration as a form of creativity support, a recurrent theme were questions about who or what is the \emph{author} of text co-written with LLMs. P8 summarised the shared fear of the ethics of \pquote{outsourcing large parts of writing to AI}. Writers shared the fundamental view that writing is a human act \nquote{for the affective experience of other humans. It feels odd to try and ask an LLM to do that}{W14}, to be \nquote{reduced to just a sort of arbiter of what is good or bad}{P13}. \nquote{This is not going to be able to mimic that authorial voice and you wouldn't really want it to anyway}{W5}.

Opinions on authorship with an AI co-writer differed, ranging from a feeling of ownership or agency due to extensive editing (W22, W25) to other writers deploring the missing \nquote{authorial voice}{W2, W5} and observing that most output was AI-generated (W20, W21, W24). W17 argued that framing writing as a \pquote{riffing} on externalized, existing knowledge, rather than an expression of internal, lived experience  (\pquote{knowledge that is outside of you and not inside of you}), reflects a \pquote{very white male way of thinking about writing}.

Several participants, including W16, had practical \pquote{concerns about the risk of other people potentially being funneled down the same generation path based on the prompts [...] is that my story or is that the other thousand people that also made the same thing?}. \nquote{You don't want the authors all over the world to just copy paste what the machine is doing}{P1}. 

\subsubsection{Concerns about cognitive deskilling.}
Historian David McCullough popularized\footnote{Earlier references include
Horatio \citep{horatio19BCars} and \nquote{Ce que l'on conçoit bien s’énonce clairement, Et les mots pour le dire arrivent aisément}{Nicolas Boileau \citep{boileau1674arts}}.}  the saying: \pquote{Writing is thinking. To write well is to think clearly. That’s why it’s so hard}. Writers expressed concerns about subcontracting out core thinking risks and eroding invaluable creative skills --- echoing recent studies by \citet{kumar2025human} or \citet{kosmyna2025your} --- and preferred for \dramabox~\nquote{to help me write as opposed to writing it in my place}{W22}.

The ease of generating scripts with AI tools, including \dramabox, was of concern (W3, W11, W21, W22, W24), as \nquote{for certain kind of writing skills, you only get better by doing it}{W17}. Writers felt that a tool that just handed writers a story removed the learning process of working on a craft: \nquote{skills get lost}{P13}, \nquote{the teaching aspect is gone}{W23}; \nquote{I'm glad this didn't exist when I was a kid because [...] I would have relied on it way too much}{W21}; \nquote{this is not allowing the learner to try by themselves. A bit like students using ChatGPT}{W24}; \nquote{aren't these tools just teaching us to become prompt writers rather than actual writer writers?}{W15}. \nquote{If [students] default to using a model like this, how are they going to have that embodied learning of working out themselves?}{W11}. W6 found \dramabox~\pquote{just too easy a tool} and preferred \pquote{students to learn how to mine themselves [...] from their own material experience}; if the tool were used in a classroom, they suggested \pquote{hiding some of those elements of the psychology, or the emotions [...] the teacher should be able to adapt what is given to the students}. A few writers disagreed on \nquote{hiding}{W6} or \nquote{handicapping}{W7} aspects of the tool in order to encourage ``creative friction'', preferring this decision to be made by the students themselves.

\subsubsection{Concerns about the automation of creative writing.}

Finally, the framing of creativity as a search process for increasing productivity worried some writers because of the risks of misuse by users who simply want to generate outputs at scale: \nquote{there is a difference between writing and creating text}{W5} and \pquote{because how the tool is set up, [W2] felt it was more for mass produced scripts}. W2 critiqued the underling productivity logic of \pquote{wanting writing [instead of] a writer}, and W9 urged us to be cautious, as \pquote{this is absolutely amazing for someone who wants to put out mediocre, sloppy content without having to pay}. Writers reiterated concerns about the proliferation of AI \emph{slop} (W3, W9) and about the automation of creative jobs \citep{weidinger2021ethical,jiang2023ai} with lower standards for creative writing (W4, W9), reducing the writer using LLMs to be \nquote{just a sort of arbiter of what is good or bad}{P13}.

\subsection{Iterative development}
\label{subsec:plan9}

Writers' concerns about cognitive deskilling prompt us to {\bf make \dramabox~more into a learning tool} for casual writers. Their desire for analytical aid prompted us to add a {\bf feedback functionality on an existing script}.

These two creativity-focused requests from writers were to allow them to rework their own material, and to make \dramabox~a tool for learning more than for generating. On the former, we added the functionality of {\bf incorporating and reworking the writer's existing material} early on; we implemented it as script uploading in order to parse it into the story plan and scene plans of the \dramabox~drama manager, and evaluated it before the first writing sessions. On the latter, we implemented a simple chatbot that has access to the script and story-plan as well as a list of narratology-focused questions, so that \dramabox~can {\bf provide feedback on the script so far via the chat interface}.

After adding all the new functionalities detailed in Table \ref{tab:decisions}, we validated our improvements during design interviews with new participants who were new to \dramabox~(interactive theatre makers P16 and P17, youth theatre producer P18). We interacted with the public-domain script of Ed Wood's \emph{Plan 9 from Outer Space}\footnote{\url{https://en.wikipedia.org/wiki/Plan_9_from_Outer_Space}}. Participants praised that the chatbot within \dramabox~can give feedback on the weakest (in a narratological sense) character in the script, suggest ways to develop it, and that we can prompt \dramabox~with that feedback to regenerate certain scenes of the script. 

\subsection{Preliminary results on using \dramabox~for encouraging creative writing among casual writers}

\begin{figure}
    \centering
    \includegraphics[width=0.49\linewidth]{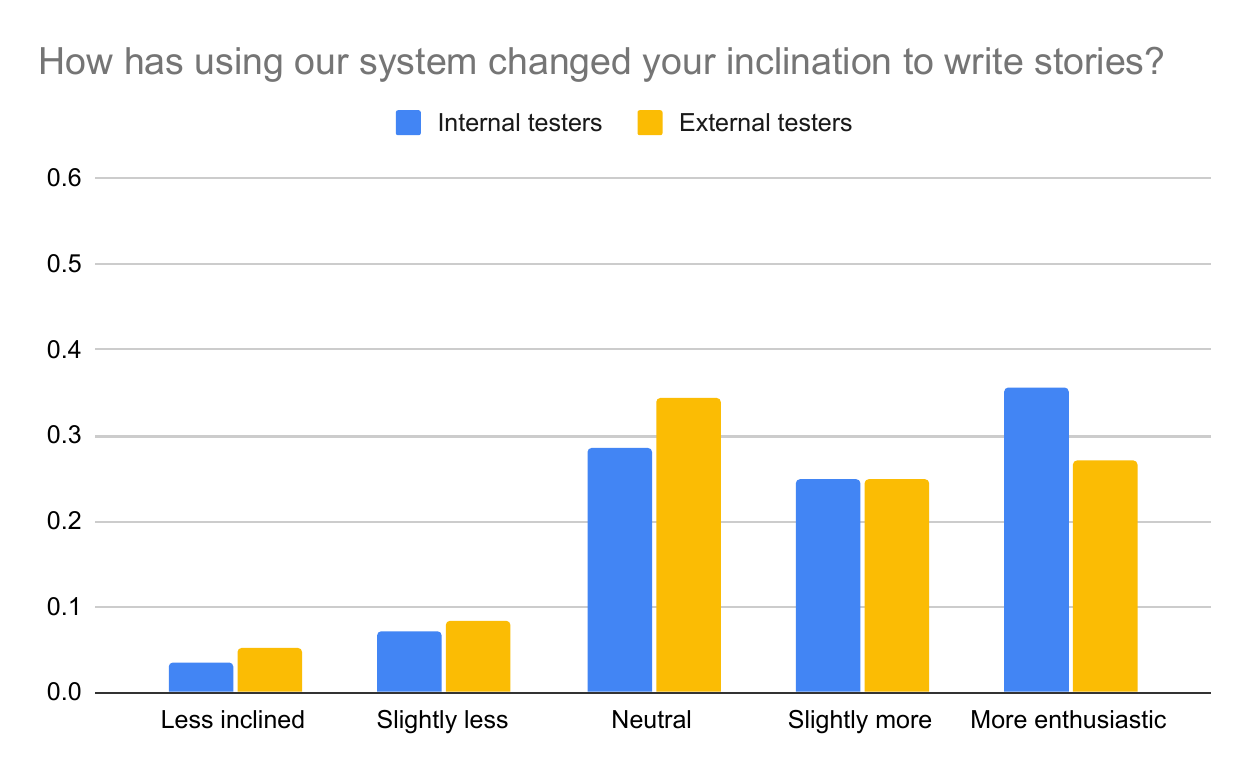}
    \includegraphics[width=0.49\linewidth]{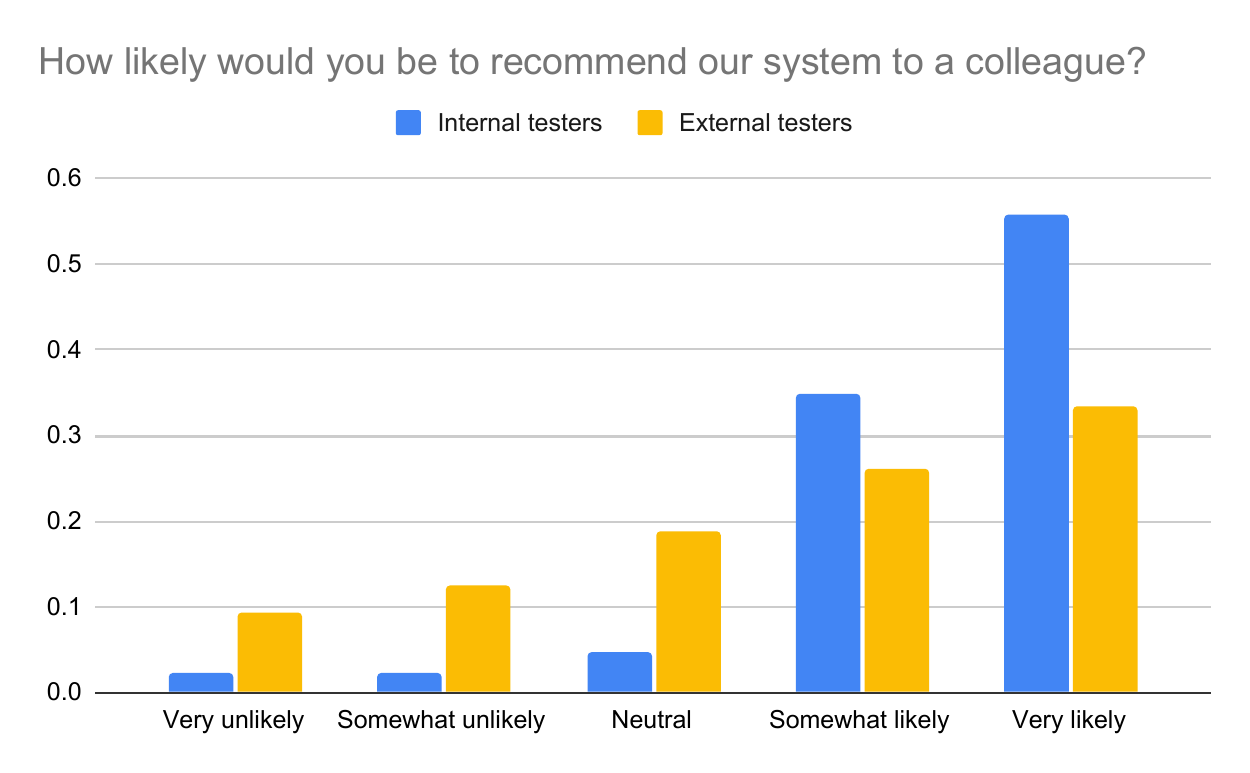}
    \caption{Inclination to write stories after using \dramabox~(left), and recommending \dramabox~to a colleague (right), as a fraction of the responses, for internal and external testers, who included a mixture of novice and experienced writers.}
    \Description{Two bar chart plots with 5 categories on the X axis and fractions on the Y axis. The titles are "How has using \dramabox~changed your inclination to write stories?" (left) and "How likely would you be to recommend \dramabox~to a colleague?" (right). Legends show three categories: internal testers, external testers.}
    \label{fig:inclination}
\end{figure}


The two groups of internal and external testers signed up voluntarily to use the app, and based on their self-description, contained a large proportion of novice and amateur writers. We are encouraged by the positive responses of the internal and external testers (including novice and experienced writers) to questions: ``How has using \dramabox~changed your inclination to write stories?'' and ``How likely would you be to recommend \dramabox~to a colleague?'' (Figure \ref{fig:inclination}).

Several of the writers, who teach drama to teenagers and university students, found potential in \dramabox~being \nquote{an interesting tool for teaching}{W11} \nquote{incredible for schools}{W9}, to \nquote{learn how to write narrative-wise}{W3}, to ask useful questions about character development and audience goals (W16, W20, W21). Future directions of research will thus include extending \dramabox~into a specific storytelling learning tutor.

%% file: paper_discussion.tex
\label{sec:discussion}

This paper has documented the process of involving different writers' communities in the design of \dramabox, as we were building the app and cycling through design interviews, internal team debugging, expert feedback or writers' sessions, and wider community testing and feedback (see Figure \ref{fig:dev-cycle}). Our findings show both disconnects and bridges between technical affordances of \dramabox~and LLMs, and creative needs.

\subsection{Summary of our findings}

The 4 previous sections present results for 4 different design choices.

In Section \ref{sec:quality}, we show that despite trying to improve the storytelling quality of the tool using self-play and self-evaluation, the writing style and the quality of the generated dialogue are seen as inferior to how \dramabox~would structure the story. In Section \ref{sec:gardeners-architects}, we discuss how the UI could cater both to Gardeners and Architects, but that writers want fine-grained control of the story and some of them prefer the option of conversing with a chatbot interface. In Section \ref{sec:hierarchy}, we probe the Western assumptions behind a hierarchy-based approach to narratology (as well as inherent LLM biases), and collect critical feedback on whether \dramabox~is capable to support various localized cultural traditions and whether it can be useful for playwriting, screenwriting or for nonlinear writer workflows. Finally, in Section \ref{sec:creativity}, we confront our vision of convergent iteration with writers' expectations of creativity support, and uncover both potential applications as a teaching tool, and concerns around creative deskilling and automation of creative writing.
As W3 commented, AI tools for creatives raise the question of ``where the boundary should sit'' and \dramabox~asked the writers what parts of writing they were comfortable delegating to tools versus what they considered central to their craft.

\subsection{Confronting the logics of the developers and of the writers}

Our process highlights tensions between logics of
scale, ease, frictionless use of technology (which we could be purposefully or inadvertently championing), and the
logics of expertise, specificity and creative control of writers (which may be difficult to satisfy).

\subsubsection{About our design choice to achieve high-quality, consistent Drama Managers (Section \ref{sec:quality}).}

Our findings revealed that what as developers we can categorize as  system optimization is often perceived by writers as technological constraint. Current AI systems are built to maximise likelihood, whereas great stories must \nquote{pierce the expectations}{P2}. Writers' requests made during design interviews or writer sessions would sometimes explicitly resist the LLMs tendency towards narrative coherence, as the writers dreamed of an \nquote{absurdity dial}{W7} that allows for the illogical, and would prefer to simulate \nquote{stream-of-consciousness}{P1} dialogue where grammar is intentionally flouted to achieve emotional power or to veer away from the expected, with the objective of highlighting messy human structures rather than smoothing them over. \nquote{You don't actually want a performance partner to be particularly solid, you want them to be fluid}{P18}.

Echoing the words of P1, there was interest in ``going beyond the human frame'': \nquote{a lot of computer scientists [...] tend to be quite interested in getting LLMs to do things that are humanlike, whereas artists tend to be much more interested in figuring out the opposite [...] the inconsistencies, the faulty things, those are the most interesting}{W14}.

\subsubsection{About our design choices of Architect views (Section \ref{sec:gardeners-architects}) that allow to work on hierarchical story structures (Section \ref{sec:hierarchy}).}

\dramabox~prioritised a structural scaffolding to make coherent stories, whereas participants had nonlinear writing workflows, and found \nquote{Save The Cat\footnote{Reference to the standard textbook used for learning screenwriting by \citet{snyder2023save}.}}{W9}, \nquote{sort of storytelling by numbers}{P4} too constraining.

\subsubsection{About the design choice of using hierarchically-structured Drama Managers with an out-of-the-box LLM (Section \ref{sec:hierarchy}).}

Writers questioned how universal can these models be, and whether they can generalize by producing truly diverse narratives. Participants identified that various LLM models often treat non-Western cultural details as superficial ``embellishments'' on top of a Western story structure, or for example, that the Western storytelling systems will bias toward interiority whereas South Asian traditions may prioritize events and folktales over character psychology.

\subsubsection{Autoethnography: constraints of participatory design with generative AI.}

Some of the writers' desiderata can contradict some of our design decisions, whereas others can be accommodated.

Much of the feedback received e.g., around ``cultural flatness'' and the treatment of identity as ornamentation, points to limitations in the underlying general-purpose LLM architectures. As argued by \citet{yang2020re}, the challenge of designing for AI is fundamentally difficult because of the uncertainty surrounding the AI's capabilities and complexity of its outputs: the AI system's ``logic'' is both probabilistic and opaque. The mismatch of AI and human capabilities reveals itself slowly: \nquote{Interacting with chatbots as a form of intelligence and a partner in discourse taught me how complicated theatrical engagements are and how much is taken for granted by the performer and also the audience}{P18}. While participatory feedback suggests a need for a completely different approach to AI design, the development team faces the constraint that changing the ``entire underlying structure'' of a massive model is often infeasible.

These conflicts are not new in technological design. \citet{suchman1987plans} sees these gaps between technical planning and the improvisational reality of human practice as a central tension in the technological design process. Such collisions suggest  that participatory AI for creativity cannot just be about fixing a tool to meet a consensus but can and should become a design process that allows these conflicting logics to coexist. Following \citet{disalvo2015adversarial} framework of Adversarial Design, we frame these collisions not as failures to be fixed, but as opportunities for debate.


\subsection{Opportunities for transforming \dramabox~into specialized tools}
\label{ref:subsec:opportunities}

The feedback we gathered for \dramabox~indicates opportunities for a fundamental shift rethinking of AI's role in creativity, which also aligns with discussions in human-centered co-creativity, where the goal is to ``support and elevate'' rather than replace human labor. One consistent feedback we got was the ask to move away from the paradigm of the story generators toward a more nuanced role as an analytical and diagnostic partner, as a world-building \emph{sandbox} or a character-drive \emph{plaground}.



\subsubsection{Using \dramabox~as an educational tool for writers and actors.}
\label{subsubsec:teaching}

Rather than focusing on the production of dialogue or complete stories, the system could move towards providing high-level diagnostic feedback \citep{rashkin2025help} as a storytelling tutor: \nquote{the most helpful output was the analysis of what I wrote}{W25}. This trajectory directly addresses anxieties regarding cognitive deskilling by effectively positioning the AI as a sidekick that highlights creative possibilities, and becomes a pedagogical tool. W7 suggested \dramabox~to produce \pquote{analyze the structure of existing piece of text}, whereas W11 imagined asking students to critique \dramabox' outputs, while P8 and P18 envisioned that the rapid generation of multiple versions of scripts according to the director's specifications could be used for actors' practice. \nquote{An actor is very used to extending themselves prosthetically through forms of technology}{P18}, and could welcome exploring potentials of technology for improving their craft.

One valid concern for the deployment of AI-based tools such as \dramabox~is the value of human peer feedback and of community-building for early-career writers: if novice writers are getting feedback from an LLM instead of fellow writers, what impacts might that have on their learning? A potential remedy is to use \dramabox~in the context of workshops, where actors or writers are working in small groups, interacting, probing and critiquing the outputs of \dramabox~or using them as drafts for readings. The lead authors of this paper trialed several such workshops with university students, as part of mixed creative computing and theatre curricula.

\subsubsection{Using \dramabox~for reworking screenplays.}
\label{subsubsec:upload}
As we discussed in Section \ref{subsec:plan9}, several screenwriters (and also a few theatre makers) were interested in such a functionality. \nquote{In television, the ability to quickly make edits and adjustments is essential. [...] Could I ingest an existing script and [...] then be making those adjustments with that existing script?}{P2}. \nquote{OK, can you change this character in this script? What consequences will it have for the beats for the other series?}{P6}. 

\subsubsection{Using \dramabox~as a tool for interactive storytelling.}
Playwrights P1, P17 and P18 wanted to employ \dramabox~to devise theatre\footnote{In devised theatre, the script originates from collaborative and improvisatory work by the performing ensemble; directors who employed devising include Jerzy Grotowski, Wlodzimierz Staniewski, Jacques Lecoq and Mike Leigh (film).} \citep{oddey2013devising} from the actors' improvisation: \nquote{live performance in particular is the ideal platform [...] real-time generation of a script between humans and AI}{P7}.
Several writers, including interactive film maker W4 and playwrights P1, P7/W7, P16 or W14, envisioned \dramabox~as a tool that allows a writer to interactively modify an existing narrative. One possibility, currently being explored by W4, is to explore all possible branches of the narrative \citep{huang2024if} with the intention of making an interactive film. Another possibility, envisioned by P1 and P17, is the use of \dramabox~in live theatrical productions, where the writer's script is uploaded to \dramabox~and where the audience can control the narrative while \dramabox~adapts the story based on the new instructions, keeping it consistent with the original world established by the writer and allowing for endless variations that adapt to the spontaneity of live performance. P17 imagined how Agosto Boal's \emph{forum theatre} \citep{boal2005legislative} (where excerpts from existing theatre scripts are reinterpreted by ``\emph{spectactors}'') could be further enriched by a tool that rewrites the script as it is being performed.

%% file: paper_conclusion.tex
We introduced \dramabox, an LLM-powered narratology tool for storytelling that allows convergent iteration on scripts. Recognizing the profound tensions between AI and human creativity, we positioned \dramabox~as a design probe to surface and interrogate the ideological conflicts between algorithmic development and artistic practice.

Through a multi-layered participatory study---comprising design interviews, intensive writing sessions, and large-scale testing---we engaged in critical dialogue with 42 writers and industry professionals from theatre, TV and cinema, as well as with several hundreds of testers. We explored the convergences and frictions between designer intent and writer needs, and integrated ongoing feedback from the writers into the app development (writers wanted agency, inspiration and creative friction, not constraints or UI-induced cognitive load). We view the very development of the tool as another form of convergent iteration, which happens via cycles of design, evaluation and improvements. 

The entire process of prototyping, interaction and critique allowed us to ``converge'' and to identify limitations both in our assumptions and in the technology, and to concentrate further development of \dramabox~into specialized tools. Ongoing work is now focused: first, on building a tool that can provide feedback on structure to a novice writer (and potentially be used a script feedback in a writers' room context for more experienced writers), and second, on experimenting with directors writing interactive stories and staging interactive productions, exploring the ``what if''.

We invite creatives to request access and to experiment with \dramabox\footnote{\dramabox~is available to writers who make an access request at \dramaboxlink. \dramabox~data governance is summarised in Appendix \ref{subsec:appendix-data}}, a prototype being shaped by the community, and thus hoping to give some agency over development back to the writers' community. As we observed in our sample, this community has diverse and nuanced views about AI, some of whom (like P1 and P18) are \pquote{not interested} in \pquote{playing the role of victim} of AI, preferring active and critical engagement, and having voice and agency in shaping these tools. We also hope our work can provide a roadmap for the human computer interaction and design community, to use AI tools and participatory design to surface---not resolve---adversarial tensions in AI design.

%% file: acknowledgements.tex
\section*{Acknowledgments}

The authors are grateful for inspirational discussions with Nick Lowe during the early design phase of this project.
The authors are grateful to Kory Mathewson for very helpful feedback on the manuscript, to
Maria Abi Raad,
Mavish Mahomed-Silva,
Emily Conn, 
Prasad Haridass, 
Nick Dietrich, 
Ed Hirst, 
Sapphire Worthington, 
Nilesh Ray, 
Abhishek Bapna,
Raia Hadsell, 
Adrian Bolton, 
Alexandre Moufarek, 
Valeria Oliveira 
for making the work and launch possible, to Bethanie Brownfield and Erin Drake Kajioka who worked on earlier versions of the tool, and to the many colleagues who provided feedback on \dramabox, in particular: 
Adam Connors, 
Emily Hutt, 
Dominic Rabiej, 
David Chess, 
Andrew Bolt, 
Artem Klimov, 
Daniel Kasenberg, 
Dylan Banarse, 
Emily Ma 
and Rowan Osmon.
Most importantly, we would like to thank the testers and writers involved in evaluating the tool, including
Annie Guo,
Arthur L Miller,
Martin Percy,
Genevieve Liveley,
Ian Garrett,
James Northcote,
Johann Smith,
Joyce Lao,
Judith Johnson,
Leonardo Castro Gonzalez,
Liam Jarvis,
Natalia Korczakowska,
Nigel Townsend,
Sabrina Morabito,
Simon Judd,
Sunil Manghani,
Torsa Ghosal,
Vanesa Kelly,
Victoriya Klipova,
Wen Mo,
and many other anonymous contributors.

\section*{Contributions}

RE implemented the first prototype of \dramabox.
RE, RKA and PM implemented the Drama Managers.
RKA, RE, JGMA and LLR implemented the auto-evaluation framework.
BW, PM, DW, JSE and EDK implemented the user interface.
RG designed the user interface. 
PM and BW prepared the product launch.
PM and RQ designed the human-computer interaction study.
PM, LS and RQ recruited the participants.
PM managed the logistics for the human-computer interaction study.
PM, RQ, RE, LS, SG, LLR and RKA ran the writer sessions and design interviews.
PM and RQ analysed the qualitative human-computer interaction study results (qualitative coding). 
SG and PM analysed the quantitative human-computer interaction study results (interaction signals from the UI). 
LR, EG and SM advised the project.
RE originated the project.
RE and PM managed the project.
PM, RE, RQ, BW and RKA wrote the paper.

%% file: appendix_ethics.tex
\label{sec:appendix-ethics}

\subsection{Ethical Approvals}
\label{subsec:appendix-approvals}
The empirical studies in this paper were approved by the ethics board appointed by our institution, who considered adverse impacts of LLMs upon participants (e.g., exposure to harmful and biased LLM outputs), the right to withdrawal without prejudice, and the compensation of participants. Participants signed consent forms before taking part in the study. Minimizing the privacy impacts upon participants was an important requirement to the research team running the study: for this reason we did not collect demographic data from the participants and only collected self-declared information about writing proficiency and experience of generative AI tools

Participants whom we recruited for design interviews and writer sessions were compensated for their time at a competitive performing industry consulting hourly rate.

At the beginning of each design interview sessions and of each writer session, the objectives of the workshop were discussed with the participants and critical opinions were actively invited.

\subsection{Data Governance}
\label{subsec:appendix-data}
We communicated to internal and external testers who used \dramabox~that they retain ownership of their written stories, including any additional text made with \dramabox. We also reassured testers that the development team does not publicise anything writers input into, or generate with, \dramabox, without the writer's express written permission, as well as about the confidentiality of data shared with the team.
\dramabox uses Gemini and Imagen models subject to the Vertex AI Data Governance policy, and we did not use any of the data resulting from \dramabox to train or build a new model, only to derive insights for the app and for this study.


%% file: appendix_user_survey.tex
\subsection{User Surveys}
\label{subsec:appendix-users}

Internal and external testers were asked these questions:
\begin{enumerate}
    \item What is the story ID? You can get it from the share ling on the left-hand side bar (e.g. "DH2cYKbA...", "I don't know", or "N/A")
    \item Do you find \dramabox~preferable to a standard chatbot interface to co-create stories? \\
    \emph{5-point Likert scale from ``I strongly prefer just using the standard chatbot interface'' to ``I strongly prefer the \dramabox~app''}.
    \item How has using \dramabox~changed your inclination to write stories? \\
    \emph{5-point Likert scale from ``It has made me less inclined'' to ``It has made me more enthusiastic''}.
    \item The \dramabox~app was not buggy/broken and worked very well. \\
    \emph{5-point Likert scale from ``Strongly disagree'' to ``Strongly agree''}.
    \item What could we do to improve \dramabox?
    \item How likely would you be to recommend \dramabox~to a friend/colleague if they had access? \\
    \emph{5-point Likert scale from ``Very unlikely'' to ``Very likely''}.
    \item Any other comments or feedback are gratefully received.
\end{enumerate}

Internal testers were asked this additional question:
\begin{enumerate}
    \item Do you like the feeling of creating the story one beat at a time (as \dramabox~does it), or would you rather the model generated the whole story at once?
    \emph{5-point Likert scale from ``I prefer the whole story to be generated at once'' to ``I strongly prefer one beat at a time''}.
\end{enumerate}

%% file: appendix_design_sessions.tex
\subsection{Interview Questions for Design Sessions}
\label{subsec:appendix-design}

\subsubsection{Questions asked to media producers}

We've been influenced by structural approaches like Robert McKee's, breaking a story down hierarchically. The overall Story is composed of Scenes, and each Scene of smaller units or Beats. Within our tool, users define what happens in each Beat, largely by answering guiding questions, and co-creating with the AI. A typical workflow goes from (a) logline and story context to (b) a story plan decomposed into scenes, then to (c) scene plans, each scene decomposed into beats, down to (d) script text?

\begin{enumerate}
    \item Understanding the role of AI and technology in the creative process.
    \begin{itemize}
        \item How has the relationship between technology and your job evolved over your career. How does AI fit in?
        \item Do you see potential for AI to support the creative processes in media production?
        \item What are the most exciting opportunities? What are the key constraints or challenges?
        \item Beyond assisting with traditional formats of screenplay, do you envision AI tools unlocking new forms of narrative or interactive experiences for television, streaming, or digital platforms?
        \item What aspects of the creative process do you feel are uniquely human and should not be given over to AI assistance, where AI might be unhelpful or even detrimental?
    \end{itemize}
    \item Evaluating \dramabox's design principles.
    \begin{itemize}
        \item Can you comment on the Story to Scenes to Beats workflow?
        \item Could you this tool be useful within the professional media production environments you work in? If yes, in what specific stages of the development or production cycle? Would this introduce new challenges?
        \item How could this AI tool easily integrate into existing workflows?
        \item What kind of agency or control do you think are most important for the writer? Where should the AI lead, and where should the user lead?
        \item Are there essential elements of storytelling you feel are difficult to capture within this framework?
        \item What would be your advice for technologists aiming to genuinely support and elevate, rather than inadvertently hinder, the art and craft of storytelling? 
    \end{itemize}
\end{enumerate}

\subsubsection{Questions asked to published writers}

\begin{enumerate}
    \item Understanding the role of AI and technology in the creative process.
    \begin{itemize}
        \item Have you experimented with AI tools in the creative or commissioning process? Why or why not?
        \item Do you see potential for AI to augment or support the creative processes of writers and creatives like you?
        \item What are the most exciting opportunities? What are the key constraints or challenges?
        \item Beyond assisting with traditional formats of screenplay, do you envision AI tools unlocking new forms of narrative or interactive experiences for television, streaming, or digital platforms?
        \item What aspects of the creative process do you feel are uniquely human and should not be given over to AI assistance, where AI might be unhelpful or even detrimental?
    \end{itemize}
    \item Evaluating \dramabox's design principles.
    \begin{itemize}
        \item (After sharing the concept for \dramabox) Can you comment on the Story to Scenes to Beats workflow?
        \item How does our hierarchical structure (from Story, to Scenes, to Beats) resonate with how you think about or build narratives? What feels intuitive? What might be missing or overly rigid?
        \item What do you think of the interaction between a writer and this AI tool, e.g. asking world-building or character-related questions?
        \item Are there essential elements of storytelling you feel are difficult to capture within this framework?
        \item What could make this tool really useful for your creative process?
        \item What would be your advice for technologists aiming to genuinely support and elevate, rather than inadvertently hinder, the art and craft of storytelling?
    \end{itemize}
\end{enumerate}

\subsubsection{Questions for creative writing instructors and narratologists}
\begin{enumerate}
    \item Understanding the role of AI and technology in the creative process.
    \begin{itemize}
        \item 
        How do you see emerging AI technologies changing how stories are created, consumed, or studied?
        \item 
        What are the most interesting opportunities that AI offers for narratology? 
        \item What are the significant concerns or limitations of AI from a narratology perspective? 
        \item What would need to happen for AI to become useful for storytelling and narrative building if at all? 
    \end{itemize}
    \item Evaluating \dramabox's design principles.
    \begin{itemize}
        \item Can you comment on the Story to Scenes to Beats workflow?
        \item From a narratology standpoint, what are the strengths and weaknesses of this hierarchical, beat-focused approach?
        \item How does this align with, or diverge from, various narrative theories or models you've studied? Are there strengths or limitations you immediately perceive?
        \item Are there different narrative structures or storytelling conventions from various global or historical traditions that you feel are particularly interesting or currently overlooked? Might it be fruitful to explore them into an AI-assisted tool like this?
        \item From your perspective, what are the absolute fundamental building blocks of a compelling story? How effectively do you think our described tool and its structure might engage with and help users develop these core components?
        \item Is there an inherent tension between such structures and creative freedom, particularly for new writers, and how might AI mediate this? 
        \item What would be your advice for technologists aiming to genuinely support and elevate, rather than inadvertently hinder, the art and craft of storytelling?
    \end{itemize}
\end{enumerate}

\subsubsection{Questions for academics and writers focused on cultural specificity/localization}
\begin{enumerate}
    \item Understanding the role of AI and technology in the creative process.
    \begin{itemize}
        \item As a writer and scholar who has bridged cultures, when we think about ``culturally localized stories'', what does it mean? Does it go beyond just direct translation? We’re particularly interested in learning how you think about the deeper, often less visible elements of storytelling that might require careful consideration when trying to create stories in and for different cultures?
        \item How important is the specific language used, including idioms, slang, and linguistic nuances?
        \item Are there certain plotlines or themes that resonate more or less with specific cultures?
        \item Are there culturally specific preferences or expectations for how a story is structured or unfolds?
        \item What are some common pitfalls in trying to localize stories?
    \end{itemize}
    \item AI and localization.
    \begin{itemize}
        \item How is generative AI impacting the question of culturally specific storytelling?
        \item Do you believe AI could have a role in preserving or revitalizing endangered cultural narratives or storytelling traditions? Anything you're particularly excited about? 
        \item Do you see a role for AI assisted narrative building that could showcase different cultural narratives and epics?
        \item What roles could exist beyond text based adaptations maybe something around interaction?
        \item Are  there different narrative structures that exist in the world that you feel are particularly interesting to explore through AI that might be  overlooked?  
        \item Do you see a role for AI assisted narrative building that could showcase different cultural narratives and epics?
    \end{itemize}
    \item Evaluating \dramabox's design principles.
    \begin{itemize}
        \item Do you foresee any inherent cultural assumptions or biases in this kind of hierarchical structure? Are we making assumptions about how a story must unfold or what constitutes a primary story structure that might not align with global traditions?
        \item We are evaluating the storylines generated  for possible conflict. Is the need for conflict a Western assumption for storytelling? 
        \item How might narrative structures, character archetypes, or even the purpose of storytelling differ across cultures in ways that \dramabox~would need to accommodate?
        \item What would be your advice for technologists aiming to genuinely support and elevate, rather than inadvertently hinder, the art and craft of storytelling?
    \end{itemize}
\end{enumerate}

%% file: appendix_creativity_survey.tex
\subsection{Writer Creativity Surveys}
\label{subsec:appendix-creativity}

\subsubsection{Your writing experience so far}
\begin{enumerate}
    \item What is your {\bf writing experience}?
    \emph{4 choices: ``Novice'', ``Amateur (including self-published)'', ``Published Writer'', ``Other (you can give more information when responding to the next question)''}.
    \item What {\bf sort of writing} you do? For example: science-fiction, children stories, screenplays, role-playing games, interactive fiction, etc. Please do not give personally identifiable information.
\end{enumerate}

\subsubsection{Your experience with AI systems for writing fiction}
\begin{enumerate}
    \item {\bf In your creative practice}, do you use AI systems to write text?
    \emph{5-point Likert scale from ``Never'' to ``Always''}.
    \item If relevant, what {\bf kind of AI systems} have you used in your creative practice?
    \item What {\bf benefits (if any)} did AI bring to your creative practice?
For example: inspiration, exploration, automation of repetitive tasks, etc.
Please do not give personally identifiable information.
    \item What adjectives might describe {\bf your experience} of AI so far?
For example: inspiring, interesting, confusing, mysterious, cumbersome, helpful, useful, concerning, unhelpful, unnecessary, etc.
    \item {\bf Outside of your creative practice}, do you use AI systems to produce content?
    \emph{5-point Likert scale from ``Never'' to ``Always''.}
\end{enumerate}

\subsubsection{Creativity Support Index of the AI writing tool}
\label{app:csi}

This section contains questions that will allow us to compute a Creativity Support Index for the AI writing tool. Most of these questions overlap with those in the previous section.

\begin{itemize}
    \item The AI system allowed other people to work with me easily. [1 to 10)
    \item It was really easy to share ideas and designs with other people inside this system or tool. (1 to 10)
    \item I would be happy to use this system or tool on a regular basis. (1 to 10)
    \item I enjoyed using the system or tool. (1 to 10)
    \item It was easy for me to explore many different ideas, options, designs, or outcomes, using this system or tool. (1 to 10)
    \item The AI system was helpful in allowing me to track different ideas, outcomes, or possibilities. (1 to 10)
    \item I was able to be very creative while doing the activity inside this system or tool. (1 to 10)
    \item The system or tool allowed me to be very expressive. (1 to 10)
    \item My attention was fully tuned to the activity, and I forgot about the AI system that I was using. (1 to 10)
    \item I became so absorbed in the activity that I forgot about the AI system that I was using. (1 to 10)
    \item What I was able to produce was worth the effort I had to exert to produce it. (1 to 10)
    \item I was satisfied with what I got out of the system or tool. (1 to 10)
\end{itemize}

In the questions above, 1 corresponds to ``Strongly disagree'' and 10 to ``Strongly agree''.

\subsubsection{Notes on the Calculation of the Creativity Support Index Score}

For the following question:
``When writing comedy material, it is most important that I'm able to'':
the participant was shown two choices of response and asked to choose one of those two choices. There were six possible responses (listed below). Given that we consider pairs of 2 different responses at a time, there are $C_6^2 = \frac{6!}{4!2!} = 15$ unique pairwise choices.

\begin{itemize}
    \item Be creative and expressive
    \item Become immersed in the activity
    \item Enjoy using the system or tool
    \item Explore many different ideas, outcomes, or possibilities
    \item Produce results that are worth the effort I put in
    \item Work with other people
\end{itemize}

All the questions listed in this section \ref{app:csi} are directly taken from \citep{cherry2014quantifying} and the NASA Task Load Index \citep{hart1988development}, and are used as is for easy of reproduction of results. Note that the questions pertaining to human-human collaboration are less pertinent to our study.

%% file: appendix_writer_sessions.tex
\subsection{Interview Questions for Writer Sessions}
\label{subsec:appendix-writers}

\subsubsection{Focus group questions}

\begin{enumerate}
    \item Did you find any of the generated outputs inspiring? If so, could you recall one output that was usable and explain in what way it helped you write? Did you find any such outputs?
    \item Did you find any of the generated outputs helpful within the professional production environments you work in? In what specific stages of the production lifecycle could such a tool potentially add the most value, or perhaps introduce new challenges?
    \item Did you find that \dramabox~could be useful for learning narrative structure or encouraging writing?
    \item Are there essential elements of storytelling you feel might be difficult to capture or encourage within \dramabox?
    \item Do you find the \dramabox~workflow easier/the same/harder than using ChatGPT or Gemini?
    \item Were any of the generated outputs that were presented culturally inappropriate in some way?
    \item Did you have any concerns about ownership or agency when generating outputs?
    \item Did you have concerns about copyright when using this tool? 
    \item What would be your advice for technologists aiming to genuinely support and elevate, rather than inadvertently hinder, the art and craft of storytelling?
\end{enumerate}